\documentclass{siamonline220329}
\usepackage{graphicx} 
\usepackage{mathtools,comment}
\usepackage{amsmath}
\usepackage{amsfonts}
\usepackage{dirtytalk}
\usepackage{caption}
\usepackage{subcaption}
\usepackage{color}
\usepackage{hyperref}
\usepackage{xcolor}

\title{Bridging the Gap between Koopmanism and Response Theory: Using Natural Variability to Predict Forced Response}

\author{Niccol\`o Zagli \thanks{Nordita, Stockholm University and KTH Royal Institute of Technology - Hannes Alfv\'ens v\"ag 12, SE-106 91 Stockholm, Sweden (\email{niccolo.zagli@su.se})} 
 \and Matthew J. Colbrook \thanks{{DAMTP, University of Cambridge, Cambridge, CB3 0WA, UK}}
 \and Valerio Lucarini \thanks{{School of Computing and Mathematical Sciences, University of Leicester, Leicester, LE17LE, UK}}
 \and Igor Mezi\'c \thanks{University of California, Santa Barbara, CA 93106, USA}
 \and John Moroney \footnotemark[3]
 }
 \headers{Bridging the Gap between Koopmanism and Response Theory}{N. Zagli, M. J. Colbrook, V. Lucarini, I. Mezi\'c, J. Moroney}

\begin{document}
\maketitle

\begin{abstract}
The fluctuation-dissipation theorem is a cornerstone result in statistical mechanics that can be used to translate the statistics of the free natural variability of a system into information on its forced response to perturbations. By combining this viewpoint on response theory with the key ingredients of Koopmanism, it is possible to deconstruct virtually any response operator into a sum of terms, each associated with a specific mode of natural variability of the system. This dramatically improves the interpretability of the resulting response formulas. We show here on three simple yet mathematically meaningful examples how to use the Extended Dynamical Mode Decomposition (EDMD) algorithm on an individual trajectory of the system to compute with high accuracy correlation functions as well as Green functions associated with acting forcings. This demonstrates the great potential of using Koopman analysis for the key problem of evaluating and testing the sensitivity of a complex system. 
\end{abstract}

\maketitle
\section{Introduction}

Despite the charming aphorism \textit{more is different} \cite{Anderson1972}, no unique definition of \textit{complexity} exists, nor does a coherent conceptual or mathematical framework to fully understand it. Nonetheless,  the past few decades have witnessed great advances in our understanding of complex systems such as ecosystems, multiagent models, spin glasses, turbulent flows, and the climate. The awarding of the 2021 Nobel Prize in physics to three scientists who have provided fundamental contributions in these research areas has largely been seen as a coronation of these scientific endeavours; see a comprehensive discussion in \cite{Gupta2021,Lovejoy2022,Ravishankara2022}.
Often complexity is associated with the presence of variability on multiple and interacting scales and noisy signals originating from an underlying deterministic chaotic dynamics or stochastic sources of randomness. The climate represents the paradigmatic example of a complex system as it is a high-dimensional, forced, dissipative, chaotic and out-of-equilibrium system featuring internal degrees of freedom acting on about 15 orders of magnitude in space and in time \cite{Ghil2020,LucariniChekroun2023}. 

The availability of ever-growing amounts of data and the rapid development of machine learning techniques are creating a paradigm shift in the modelling of complex systems: it has been recognized that top-down, theory-driven approaches need to be blended with bottom-up, data-driven methods. 
Data-centric approaches have indeed had huge success in multiple research areas \cite{Raissi2019,Pershin2022}. 

A powerful synthesis between data-driven and theory-informed approaches can be found \cite{Mezic2005,Mezic2020AMS}. \textit{Koopmanism} is a theoretical framework whereby the evolution of a nonlinear system can be reconstructed by studying the properties of a linear operator - the Koopman operator - acting on observables of the system \cite{Budisic2012}. This is a powerful tool for analysing the dynamics of the system and for obtaining statistical information about it. Koopmanism has found applications in areas such as fluid dynamics, control theory, data analysis, and many others \cite{Budisic2012,Brunton2022}. The power and limits of Koopman learning of dynamical systems from data can be put into rigorous mathematical foundations, see \cite{colbrook2024}. The standard algorithm used to approximate the Koopman operator from data is represented by the Dynamic Mode Decomposition (DMD) method  \cite{DMD, rowley_mezi_bagheri_schlatter_henningson_2009, KutzBrunton2016}. DMD is computationally efficient and provides a scalable dimensionality reduction for high-dimensional data. DMD is valid for both experimental and simulated data, as it is based entirely on measurement data and does not require knowledge of the governing equations \cite{Brunton2022}. 

A recent generalization of DMD is Extended DMD (EDMD) ~\cite{WilliamsKevrekedis2015,LDBK2017}. EDMD's improvement upon DMD comes from the flexibility in choosing the dictionary of observables allowing for estimating and approximating the  Koopman operator. Variants and extensions of EDMD have been developed to investigate high-dimensional systems and to control infinite dimensional projection errors, see ~\cite{Colbrook2023_MultiverseDMD} for a comprehensive review.

\subsection{Linking Forced and Free Fluctuations}
Predicting the response of a complex system to perturbations and identifying possible critical behaviour associated with diverging sensitivity and slow decay of correlations is of extreme importance. Efficient prediction requires the possibility of using information gathered on the system in the absence of the acting perturbation. For systems near equilibrium, the theoretical foundation for such an operation is provided by the fluctuation-dissipation theorem (FDT) \cite{Kubo1966}, which allows one to write a (linear) response operator for an observable of interest as a rather intuitive correlation between such observable and a suitably defined conjugated variable. The FDT can be extended to stochastic systems far from equilibrium \cite{MArconi2008,Sarracino2019} and in such a way that nonlinear effects can be accounted for \cite{LC12}. Within a stochastic framework, it is possible to show that linear response theory is valid under rather general conditions and applicable response formulas can be derived \cite{Hairer2010}. Linking free and forced variability is harder in the case of dissipative chaotic systems: in this case the classical form of FDT is not valid, even if it is possible to (painfully) construct a rigorous linear response theory \cite{Ruelle1997DifferentiationSRBstates,liverani2006,Ruelle2009,B14}; see discussion and applications in \cite{LucariniSarno2011,Lucarini2017}. 

Complex systems can feature either smooth response to perturbations or tipping behaviour due to critical transitions. Criticality is often associated with cataclysmic and undesirable events ~\cite{SchefferTipping,Sornette2004} and with the presence of multistable behaviour \cite{FeudeletAl2018FocusIssue,Ashwin2012}.
In the smooth response regime, one can use response formulas to predict the change in the statistics due to the applied perturbations \cite{Ruelle2009,Baiesi2013,Lucarini2017}. Critical transitions \cite{SchefferTipping,KUEHN20111020} are intimately related to the occurrence of high sensitivity of the system to perturbations: the response operators diverge exactly because the system decorrelates very slowly \cite{ChekrounJSPI,GutierrezLucarini2022,FirstPaper,ZagliPavliotisLucarini2023,Zagli_2024}. 
The fundamental equivalence between high sensitivity to perturbations and slow decay of correlations is the basis of the so-called early warning indicators (EWIs) \cite{Lenton2012}.  
Whilst the FDT allows one to reconstruct forced variability from the observations of the natural fluctuations of the system, it falls short in terms of interpretability. Indeed, it does not include provisions for decomposing the response of the system to perturbations into a sum of terms, each associated with a mode of natural variability of the system, as one instead obtains, for example, when looking at the linear optical response of matter to an electromagnetic field \cite{CohenT}. 

Building on \cite{ChekrounJSPI}, Ref. \cite{GutierrezLucarini2022} proposed a new expression of the response operators based on a spectral decomposition based on the eigenvalues/eigenvectors of the Kolmogorov operator ~\cite{pavliotisbook2014}, which is the reformulation of the Koopman operator in the context of stochastic dynamics generated by Stochastic Differential Equations (SDEs) \cite{Zic2020}.  We refer the reader to ~\cite{Mezic2004} for the definition of the stochastic Koopman operator for more general stochastic systems. We also mention previous works on response theory from a transfer operator perspective in both deterministic and stochastic settings \cite{Lucarini2016,Antown2022,Chandramoorthy2022}.

Such a formal reconstruction of the response in terms of specific modes of unperturbed variability allows for identifying feedbacks across scales in complex systems and is especially promising in the context of climate research \cite{LucariniChekroun2023}: it explains the mathematical mechanism behind the occurrence of tipping points \cite{Tantet_2018}, clarifies an ansatz by Hasselman and collaborators \cite{Hasselmann1993}, and supports recently proposed data-driven methods for studying the climate response \cite{Torres2021a,Bastiaansen2021}.
\subsection{This Work}

We show that it is possible to use the data-driven methods aimed at accurately computing the eigenvalues and eigenvectors of the Koopman/Kolmogorov operator, and specifically EDMD, to compute with a high degree of accuracy the operators describing the response of such systems to perturbations. Hence, our goal is to show that the Koopmanism viewpoint on dynamics can be augmented in such a way that it wields the potential to predict the sensitivity of a system to general perturbations. This has several key implications in terms of assessing the resilience of a system and in particular its proximity to critical behaviour.  

The paper is structured as follows. Section \ref{sec: generalities} provides a succinct yet self-contained mathematical account of how the operator formalism used here sheds new light on the interpretation of the key results of response theory.  Section \ref{spectralreconsruction} translates the key results of section \ref{sec: generalities} into the EDMD language, thus providing formulas that can be readily applied in numerical investigations.  Section \ref{sec: numerical experiments} discusses the results of the three mathematical models studied here. The first model is a one-dimensional chaotic map introduced in \cite{Wormell2025_EDMD}. The second model is the stochastically forced version of the two-dimensional chaotic map introduced in \cite{Slipantschuk2020}. The latter is a variant of the classical Arnold's cat map, where dissipation is introduced so that the invariant measure is singular with respect to Lebesgue. The study of this model in the weak-noise limit is particularly interesting. The third model is a stochastically forced two-dimensional gradient flow associated with a double-well potential. The EDMD analysis of this classical equilibrium system has been showcased in \cite{Schutte2016}. In section \ref{conclusions}, we present our conclusions and perspectives for future work.

\section{Statistical Properties of stochastic systems: a Markov semigroup approach}
\label{sec: generalities}
We consider an autonomous system described by a set of SDEs 
\begin{equation}
\label{eq: Stochastic Differential equations}
    \mathrm{d}\mathbf{x} = \mathbf{F}(\mathbf{x}) \mathrm{d}t + \sigma \boldsymbol{\Sigma}(\mathbf{x}) \mathrm{d}\mathbf{W}_t , \quad \mathbf{x} \in \mathcal{M} \subseteq \mathbb{R}^d
\end{equation}
where the, generally non-linear, vector field $\mathbf{F}(\mathbf{x}):\mathcal{M} \to \mathcal{M}$ represents the drift part of the dynamics of the system,  $\mathbf{\Sigma}(\mathbf{x}) : \mathcal{M} \to \text{Mat}_{\mathbb{R}}\left( d \times q \right)$ is a state-dependent volatility matrix shaping the stochastic part of the evolution, $\sigma >0$ determines the strength of the noise term, and $\mathbf{W}_t = (W^1_t, \dots , W^q_t)$  is a $\mathbb{R}^q$-valued Wiener process,  $q$ not necessarily equal to $d$, with mutually independent components. In this paper, we consider settings where $\mathcal{M}$ is $\mathbb{R}^d$ or a periodic domain of size L in each direction, $\mathbb{T}_L^d = \mathbb{R}^d/(L\mathbb{Z})^d$. It is well known that the statistical properties of the stochastic trajectories generated by the stochastic process determined by  \eqref{eq: Stochastic Differential equations} are encoded in the probability density $\rho(\mathbf{x},t)$ solution of the Fokker-Planck (or Forward Kolmogorov) equation 
\begin{equation}
\label{eq: Fokker Planck}
    \frac{\partial \rho}{\partial t} = \mathcal{L}_0 \rho \coloneqq - \nabla \cdot \left( \mathbf{F}\rho\right) + \frac{\sigma^2}{2} \mathbf{D}^2 : \left( \boldsymbol{\Sigma}\boldsymbol{\Sigma}^T\rho \right),
\end{equation}
where $\mathbf{D}_{ij}= \frac{\partial^2}{\partial x_i \partial x_j}$ is the matrix of second derivatives.
In general, the above equation should be interpreted in a weak sense (in the sense of distributions). However, the smoothness and regularity of the probability density $\rho(\mathbf{x},t)$ are, in most stochastic settings, guaranteed as the noise yields a regularising effect provided that it can sufficiently spread over all the phase space. Such physical intuition can be rigorously justified by requiring the Fokker-Planck operator $\mathcal{L}_0$ to be hypoelliptic, see \cite[Ch. 6]{pavliotisbook2014}. The assumption of hypoellipticity is quite general and can encompass situations featuring degenerate noise acting only on part of the systems' equations, which are often encountered in applications. Hypoellipticity, together with other mild regularity and growth conditions on drift and coefficient terms of equation \eqref{eq: Stochastic Differential equations}, see \cite[Ch. 4]{pavliotisbook2014}, guarantees the existence of a unique invariant measure $\mu_0$, solution of the stationary Fokker Planck equation, which is absolutely continuous with respect to Lebesgue; that is, it admits a smooth probability density $\rho_0(\mathbf{x})$ such that $\mathrm{d}\mu_0(\mathbf{x})= \rho_0(\mathbf{x}) \mathrm{d}\mathbf{x}$. Furthermore, we consider settings in which the invariant measure is ergodic, so that $\mu_0$ provides information on the expectation value at stationarity of any observable $f$ as
\begin{equation}
\langle f \rangle_{\mu_0} = \int_\mathcal{M} f(\mathbf{x})\mathrm{d}\mu_0(\mathbf{x}) =  \lim_{T \to + \infty}\frac{1}{T}\int_0^T f(\mathbf{x}(t)) \mathrm{d}t.
\end{equation}
where $\mathbf{x}(t)$ is a trajectory generated by the set of SDEs \eqref{eq: Stochastic Differential equations} and the last equality holds because of ergodicity. We refer the reader to \cite[Ch. 2]{pavliotis2014book} and references therein for conditions on the drift and diffusion terms that lead to an ergodic stochastic process.
\\
It turns out that many other statistical properties of the stochastic system  \eqref{eq: Stochastic Differential equations} can be investigated through a functional analytical operator theory. This alternative description focuses on the time-evolution of (statistics of) observables $f$ of the system rather than individual stochastic trajectories.  In particular, any autonomous (time-homogenous) Markov stochastic process can be described in terms of a semigroup of operators, i.e., a family $\mathbf{P} = \left( P_t \right)_{t \geq 0}$ of linear operators parametrised by the temporal variable $t$. The semigroup $\mathbf{P}$ is defined, at each point in time, through its action on bounded observables $f$ of the system as
\begin{equation}
\label{eq: Def Markov Semigroup}
	\left( P_t f \right)(\mathbf{x})= \int_{\mathcal{M}} f(\mathbf{y}) p(\mathbf{y},t|\mathbf{x})\mathrm{d}\mathbf{y},
\end{equation}
where the $ p(\mathbf{y},t|\mathbf{x})$ represents the transition probability of getting to point $\mathbf{y}$ at time $t$ from point $\mathbf{x}$ at time $t=0$.  A more general definition of the stochastic Koopman operator encompassing a broader class of dynamical processes can be found in \cite{Mezic2004,Mezic2005}. If the system is purely deterministic, its evolution is described in terms of a flow $S^t$ uniquely mapping initial conditions $\mathbf{x}_0$ to their time evolved state $\mathbf{x}(t) = S^t \mathbf{x}_0$. In this case, the transition probabilities are given by the deterministic flow as $ p(\mathbf{y},t|\mathbf{x}) = \delta(\mathbf{y} - S^t\mathbf{x})$ and the semigroup is given by
\begin{equation}
	\left( P^{det}_t f \right)(\mathbf{x})= \int_{\mathcal{M}} f(\mathbf{y})  \delta\left(\mathbf{y} - S^t\mathbf{x}\right)\mathrm{d}\mathbf{y} = f(S^t \mathbf{x}),
\end{equation}
which corresponds to the usual definition of the Koopman group in deterministic settings generated by Lipschitz Ordinary Differential Equations ~\cite{lasota,Budisic2012,Brunton2022}. The previous equation shows that $P^{det}_t$ determines the evolution in time of observables $f$ of the system. A similar interpretation holds for $P_t$ with the caveat that its action should be interpreted in a statistical sense. Equation \eqref{eq: Def Markov Semigroup} represents an expectation value over all final states $\mathbf{y}$ weighted according to the transition probabilities $p(\mathbf{y},t|\mathbf{x})$. It is a classical result that the existence of the unique invariant measure $\mu_0$ implies that the semigroup $\mathbf{P}$, when acting on functions $f \in L^2_{\mu_0} = \{f : \int f^2(\mathbf{x})\mathrm{d}\mu_0( \mathbf{x}) < \infty\}$, is a strongly continuous semigroup ~\cite{Bakry2014} - see ~\cite{engelsemigroups2006,EngelNagel2000} for an introduction to semigroup theory. This means that it is possible to define an operator  $\mathcal{K}_0$ such that the action of the semigroup can be written, at every time $t$,  in an exponential form as
\begin{equation}
\label{eq: generator}
    P_t = e^{t \mathcal{K}_0}
\end{equation}
The operator $\mathcal{K}_0$ is called the generator of the Markov semigroup, it is (usually) an unbounded operator and its domain $\mathcal{D}(\mathcal{K}_0)$ is dense in $L^2_{\mu_0}$. The action of the semigroup $\mathbf{P}$ is then uniquely defined by its generator $\mathcal{K}_0$ (and its domain). According to the definitions \eqref{eq: Def Markov Semigroup} and \eqref{eq: generator}, $\mathcal{K}_0$ represents the generator of time translation of observables of the system. More specifically, given an observable $f$ of the system, the evolution in time of its expectation $\bar f(x,t) = \left(P_t f\right)\left( \mathbf{x}\right) $ is determined by 
\begin{equation}
\frac{\partial \bar f}{\partial t} = \mathcal{K}_0\bar{f}.
\end{equation}
The above equation is known as the Backward Kolmogorov equation and it determines the evolution of expectations of observables as an initial value problem together with the initial condition $\bar f(x,0) = \left(P_0 f\right)\left( \mathbf{x}\right)  = f(x)$. For a stochastic Markov process that does not feature jumps, such as the one generated by \eqref{eq: Stochastic Differential equations}, the generator $\mathcal{K}_0$ takes the name of Kolmogorov operator and can be identified as the second order differential operator given by \cite{pavliotisbook2014}
\begin{equation}
\mathcal{K}_0  =  \mathbf{F}(\mathbf{x})\cdot \nabla  + \frac{\sigma^2}{2} \left( \boldsymbol{\Sigma}\boldsymbol{\Sigma}^T  \right): \mathbf{D}^2.
\end{equation}
In the case of a jump-diffusive process, a generalisation of the above operator, featuring a non-local term, can be defined. We refer the reader to ~\cite{Chekroun2024Kolmogorov} for a Kolmogorov analysis of chaotic jump-diffusive processes.
Interestingly, many physically relevant statistical properties of the stochastic process \eqref{eq: Stochastic Differential equations} can be characterised in terms of the spectral features of the generator $\mathcal{K}_0$ as described in the next section. 
\subsection{Stochastic resonances, Correlation functions and Response Operators}
A quantity of interest for a statistical mechanical description of nonequilibrium systems is represented by the correlation function $C_{fg}(t)$ between two observables $f,g \in \mathbb{C}$ 
\begin{equation}
\label{eq: correlation function}
    C_{f,g}(t) \coloneqq \int_{\mathcal{M}} \int_{\mathcal{M}}  f(\mathbf{y}) g^*(\mathbf{x}) p(\mathbf{y},t|\mathbf{x})\mathrm{d}\mu(\mathbf{x}) = \int_{\mathcal{M}} e^{\mathcal{K}t}f(\mathbf{x})g^*(\mathbf{x})\mathrm{d}\mu(\mathbf{x}) ,
\end{equation}
where, without loss of generality, $f$ and $g$ have been assumed to have vanishing mean $\langle f \rangle_{\mu_0}  = \langle g \rangle_{\mu_0} = 0$ and in the last equality we have used the definition of Markov semigroup \eqref{eq: Def Markov Semigroup} and its generator $\mathcal{K}_0$. Correlation functions provide a measure of the relationship between two observables as one evolves with time. It is thus not a surprise that $C_{f,g}(t)$ not only depends on the invariant measure but also on the transition probabilities characterising the stochastic process. 
\\
More importantly, a fundamental question is to understand how the physical properties of systems change as external, time-dependent forcings push them away from their stationary state. Linear response theory provides a powerful framework to predict the forced evolution of systems in terms of response operators, encoding, in a linear regime, all positive and negative feedbacks of the system. We here consider the situation where the drift of the unperturbed system \eqref{eq: Stochastic Differential equations} is perturbed as
\begin{equation}
\label{eq: Perturbed SDEs}
    \mathrm{d}\mathbf{x} = \left( \mathbf{F}(\mathbf{x}) + \varepsilon  \mathbf{X}(\mathbf{x})T(t) \right) \mathrm{d}t + \sigma \boldsymbol{\Sigma}(\mathbf{x}) \mathrm{d}\mathbf{W}_t, 
\end{equation}
where $\varepsilon$ is a small parameter, $T(t)$ is a bounded function providing the time modulation of the perturbation and $\mathbf{X}(\mathbf{x})$ its phase space profile. The Fokker-Planck equation associated with \eqref{eq: Perturbed SDEs} is now 
 \begin{align}
 \label{eq: Perturbed FP}
&\frac{\partial \rho}{\partial t} = \big(  \mathcal{L}_0  + \varepsilon T(t) \mathcal{L}_p \big) \rho, \\
& \rho(\mathbf{x},0) = \rho_0(\mathbf{x}),
\end{align}
where we have introduced the perturbation operator $\mathcal{L}_p  \cdot  = - \nabla \cdot \left ( \mathbf{X}(\mathbf{x}) \! \! \! \! \quad \cdot  \! \! \! \! \quad   \right)$ and the choice of the initial condition reflects the fact that the perturbation starts acting on the system once it has already reached its statistical steady state. As the system is forced to operate in non-autonomous conditions, no invariant measure associated with \eqref{eq: Perturbed FP} generally exists. However, in a linear regime, it is possible to show that the expectation value of a generic observable $f$ is given by
\begin{equation}
\langle f \rangle_\mu(t) = \langle f \rangle_{\mu_0} + \varepsilon \delta^{(1)}[f](t) + o(\varepsilon).
\end{equation}
Linear response theory provides useful formulas to evaluate the first order correction as \cite{GutierrezLucarini2022}
\begin{align}
 \label{eq: linear correction}
\delta^{(1)}[f](t) &= \left( G_{f} \star T \right)\left( t \right), \quad \text{with} \\
\label{eq: Def Green Function}
 G_{f}(t) &=  \Theta(t) \int \frac{\mathcal{L}_p \rho_0}{\rho_0} e^{t \mathcal{K}_0} f \mathrm{d}\mu_0(\mathbf{x}),
\end{align}
where $\Theta(t)$ represents the Heaviside function. $G_f(t)$ is the Green's function of the system and determines the causal response of the system for any forcing $T(t)$. It depends both on the observable $f$ under consideration and the phase space dependent forcing $\mathbf{X}(\mathbf{x})$ through the perturbation operator $\mathcal{L}_p$. Equation \eqref{eq: Def Green Function} represents a quite general form of the FDT establishing a relation between the forced variability of the system prescribed by the Green's function and the statistical properties of the unperturbed system. More specifically the Green's function can be written as a suitable unperturbed correlation function 
\begin{equation}
\label{eq: Green Function as correlation}
G_f(t) = \Theta(t) C_{f , \Gamma}(t) , \quad \text{with } \Gamma(\mathbf{x}) = \frac{\mathcal{L}_p \rho_0}{\rho_0}.
\end{equation}
Contrary to equilibrium statistical mechanics, there are no first principles determining the functional form of the invariant distribution $\rho_0(\mathbf{x})$ of nonequilibrium systems, which generally depends heavily on the detailed dynamical properties of the system. As such, the observable $\Gamma$ is not known a priori for systems far from equilibrium and its physical interpretation is not generally clear. Nevertheless, equations \eqref{eq: linear correction} and \eqref{eq: Green Function as correlation} provide a solid theoretical framework to estimate response properties from the observation of the statistical dynamical behaviour of the unperturbed system. We observe that, in order for $C_{f,\Gamma}$ to be well-defined, the response observable $\Gamma$ should be in $L^2_{\mu_0}$ which requires regularity and growth conditions on both the Markov semigroup and the perturbation field. In this paper, as is typically done, we assume that the FDT \eqref{eq: Green Function as correlation} is well defined.
\\\\
It is possible to obtain a decomposition of both correlation and Green's functions in terms of the spectral properties of $\mathcal{K}_0$ in $L^2_{\mu_0}$ ~\cite{ChekrounJSPI,GutierrezLucarini2022}. Here, we further require that the stochastic dynamics \eqref{eq: Stochastic Differential equations} is associated with a quasi-compact Markov semigroup $\mathbf{P}$ such that its discrete eigenvalues whose real part are larger than $e^{- |r_{ess}|t}$ are discrete and have finite algebraic multiplicity. Here, $r_{ess} < 0$ is the essential growth bound associated with the semigroup \footnote{The essential growth bound of a semigroup $P_t$ is defined as $r_{ess} \coloneqq \inf_{t > 0 } \frac{1}{t}\ln || P_t||_{ess}$ where $|| \cdot ||_{ess}$ is the norm measuring the distance of $P_t$ from the space of compact operators using the operator norm. See \cite[Ch 4]{EngelNagel2000} for more details and \cite[Ch 5]{EngelNagel2000}  for the characterisation of quasi compact semigroups with $r_{ess}$}. Such eigenvalues of the semigroup are related to the eigenvalues of the generator through the spectral mapping theorem \cite{EngelNagel2000}, meaning that if $\lambda_j$ is an eigenvalue of $\mathcal{K}_0$ corresponding to the eigenfunction $\varphi_j(\mathbf{x})$, so is $e^{\lambda_j t}$ of $P_t=e^{t\mathcal{K}_0}$ relative to the same eigenfunction.  
The discrete, isolated, eigenvalues $\{ \lambda_j \}_{j=1}^N$, with $N$ possibly infinite, are sometimes called (stochastic) Ruelle-Pollicott resonances \cite{ChekrounJSPI} and provide a physically relevant decomposition of the action of the semigroup as, see Theorem 3.7 in \cite[Ch. 5]{EngelNagel2000}
\begin{equation}
\label{eq: spectral decomposition semigroup}
    e^{t\mathcal{K}_0} = \sum_{j=0}^N e^{\lambda_j t} \Pi_j + \mathcal{R}(t),
\end{equation}
where $\Pi_j$ represents the projection operator onto the eigenspace associated to $\lambda_j$ and $\mathcal{R}(t)$ denotes the contribution of the essential spectrum. When writing \eqref{eq: spectral decomposition semigroup}, we have assumed that the eigenvalues $\lambda_j$ are not degenerate, but a similar result holds even when their algebraic multiplicity is greater than one ~\cite{ChekrounJSPI,GutierrezLucarini2022}. The assumption of quasi-compactness of the semigroup guarantees that the effect of $\mathcal{R}(t)$ becomes progressively more negligible as $t \to \infty$ and that appropriate bounds on its norm can be found \cite{ChekrounJSPI}. In more practical terms, the condition of quasi-compactness paves the way for performing a model reduction by focusing on the contributions associated with the Ruelle-Pollicott resonances. Additionally, the smaller the essential growth bound $r_{ess}$, the further the essential spectrum is from the imaginary axis and thus less relevant the contribution of $\mathcal{R}(t)$ is.  

The eigenvalues $\lambda_j$s $ \in \mathbb{C}$, which for simplicity we consider to be ordered in lexicographic order as $\lambda_0 = 0 > \mathbf{Re}\lambda_1 \geq \mathbf{Re}\lambda_2 \geq \dots $,  all have negative real part and represent intrinsic relaxation rates of the system shaping the statistical dynamical properties of the system. An important remark is needed at this point. When considering purely deterministic chaotic systems the spectral decomposition \eqref{eq: spectral decomposition semigroup} of the Markov semigroup in $L^2_{\mu_0}$ is no longer valid. In fact, the (now) Koopman operator $\mathcal{K}_0$ is unitary in $L^2_{\mu_0}$, which, in physical terms, means that the dynamics has already settled onto the chaotic attractor and no decay towards it can be observed ~\cite{Budisic2012}.  It is known, at least for a class of chaotic systems, that in order to obtain eigenvalues bearing a physical value, such as expressing mixing rates (resonances) of the system, the operator $\mathcal{K}_0$ should be considered as acting in non-standard functional spaces adapted to the specific chaotic dynamics and its underlying geometrical structure ~\cite{liverani2006,B14}. 

Considering \eqref{eq: spectral decomposition semigroup}, neglecting the term $\mathcal{R}(t)$, and recalling the definition of correlation functions and Green's functions, it is possible to obtain the spectral decomposition
\begin{align}
\label{eq: spectral decomposition corr and Green's functions}
    C_{f,g}(t) &= \sum_{j=1}^N \alpha_j(f,g)e^{\lambda_j t}, \\ 
    G_{f}(t) &= \Theta(t) \sum_{j=1}^N \beta_j(f,\Gamma)  e^{\lambda_j t}.
\end{align}
The above equations show that these time-dependent statistical properties can be decomposed as a sum of exponentially decaying terms, whose rates are given by the eigenvalues of the Kolmogorov operator $\mathcal{K}_0$ and are thus an intrinsic property of the system. On the contrary, the parameters $\alpha_j(f,g)$ and $\beta_j(f,\Gamma)$ represent the projection of the observables of interest onto the Kolmogorov modes and are thus an observable-dependent quantity. The summation in \eqref{eq: spectral decomposition corr and Green's functions} does not include the static eigenmode associated with $\lambda_0 = 0$ as this mode provides information on the expectation values of observables, which vanishes by assumption for the observables $f$ and $g$ in $C_{fg}(t)$ and by definition for the observable $\Gamma$ in $G_{f}(t)$. The eigenvalues $\lambda_j$ generally feature an imaginary part corresponding to an oscillatory behaviour of correlation functions. This is a signature of the nonequilibrium nature of the system and it is a typical situation observed in chaotic systems. On the contrary, physical systems satisfying detailed balance are associated with Kolmogorov operators that are self-adjoint in $L^2_{\mu_0}$ and are characterised by purely real eigenvalues and absences of oscillations \cite{pavliotisbook2014}, see section \ref{sec: double well}. The dominant contribution at large times will be given by the Kolmogorov mode associated with the decay rate $\gamma = - \mathbf{Re}\lambda_1$. The quantity $\gamma$ is commonly known as the spectral gap and its positivity guarantees the exponentially fast decay of the correlation functions  \cite{B00, pavliotisbook2014, ChekrounJSPI}. However, we observe that the number $N$ of resonances $\lambda_j$ is not necessarily finite and pathological situations may occur when they accumulate towards the eigenvalue $\lambda_0 = 0$. Such situations are associated with sub-exponential decay of correlations \cite{Ruelle1986} and a signal of the dynamical system approaching a critical setting, such as the deterministic pitchfork bifurcation ~\cite{Gaspard1995}, the critical transition between competing chaotic attractors in an atmospheric circulation model \cite{Tantet_2018} and the rough parameter dependence of statistics in geophysical fluids ~\cite{Chekroun2014}. In the next section, we show how to estimate from data the hierarchy of Kolmogorov modes and how to reconstruct correlation and Green's functions using the spectral decomposition \eqref{eq: spectral decomposition corr and Green's functions}.

\section{Spectral reconstruction of Correlation and Response Operators from data }\label{spectralreconsruction}
The analytical evaluation of resonances and other spectral properties of transfer operators is quite a challenging task, especially in the settings of stochastic chaotic systems for which the Kolmogorov operator $\mathcal{K}_0$ is far from being self-adjoint, see for example ~\cite{Davies2007}. Even if possible, such analysis would require the knowledge of the drift and diffusion terms determining the evolution of the system. Numerical methods are then used to estimate the resonances from data by evaluating the spectrum of transfer matrices resulting from a Galerkin projection of the action of the semigroup onto a finite number of basis functions. A classical approach in the literature is Ulam's method where transition matrices are evaluated by first seeking a partition of the phase space into a finite number of disjointed boxes. Here, we employ a similar in spirit but more general algorithm, known as the Extended Dynamic Mode Decomposition ~\cite{WilliamsKevrekedis2015}, which allows for a Galerkin projection of the action of the semigroup onto any set of basis functions. If indicator functions are used as basis functions the Extended Dynamic Mode Decomposition (EDMD) algorithm is equivalent to Ulam's method ~\cite{Schutte2016}. However, when the eigenfunctions of the Kolmogorov operator $\mathcal{K}_0$ are smooth, EDMD is computationally less demanding than Ulam's method. We also remark that variants and extensions of EDMD have been developed to investigate high-dimensional systems, see ~\cite{Colbrook2023_MultiverseDMD} for a comprehensive review.

\subsection{The Extended Dynamic Mode Decomposition algorithm}
We briefly recall the fundamentals of the EDMD algorithm, adapted to the settings under consideration in this paper. We consider a trajectory $\mathbf{x}_i \coloneqq \mathbf{x}(t = i \Delta t)$, sampled at a fixed time-step $\Delta t$, stemming from equations \eqref{eq: Stochastic Differential equations} at stationarity. In other words, we fix  $t = 0 $ to be the time at which the system is described by the invariant measure $\mu_0$. EDMD requires as input a set of snapshot data $\mathbb{D} = \{ \left( \mathbf{x}_i, \mathbf{y}_i = \mathbf{x}_{i+1} \right)\}_{i=1}^M$. When the data is sampled from a trajectory (ergodic sampling), EDMD approximates the spectral properties of the Kolmogorov operator in $L^2_{\mu_0}$ ~\cite{WilliamsKevrekedis2015}, which are at the basis of the spectral decomposition \eqref{eq: spectral decomposition corr and Green's functions}. 
\\
We define a set of dictionary functions $\{\psi_j \}_{j=1}^N$ spanning an $N-$dimensional subspace $\mathcal{F}_N \subseteq L^2_{\mu_0}$ and we evaluate the action of the Markov semigroup on $\mathcal{F}_N$. It is customary to arrange the dictionary of variables as a vector-valued function $\boldsymbol{\Psi}: \mathcal{M} \to \mathbb{C}^{1\times N}$ 
\begin{equation}
    \boldsymbol{\Psi}(\mathbf{x}) = 
    \begin{pmatrix}
        \psi_1(\mathbf{x}) , \psi_2(\mathbf{x}) , \dots , \psi_N(\mathbf{x})
    \end{pmatrix}.
\end{equation}
EDMD provides the best least square approximation of the representation matrix of the Kolmogorov operator in the finite dimensional subspace $\mathcal{F}_N$ as
\begin{equation}
\label{eq: Kolmogorov matrix}
    \mathbf{K} = \mathbf{G}^+\mathbf{A} \in \mathbb{C}^{N\times N},
\end{equation}
where $\mathbf{G}^+$ denotes the pseudo-inverse of the matrix $\mathbf{G}$ and
\begin{align}
    \mathbf{G} &= \frac{1}{M}\sum_{i=1}^M \boldsymbol{\Psi}^*(\mathbf{x}_i) \boldsymbol{\Psi}(\mathbf{x_{i}}),\label{matrixGplus}
    \\
  \mathbf{A} &= \frac{1}{M}\sum_{i=1}^M \boldsymbol{\Psi}^*(\mathbf{x}_i) \boldsymbol{\Psi}(\mathbf{y_{i}}).\label{matrixA}
\end{align}
The spectral properties of the matrix $\mathbf{K}$ provide information on the eigenvalues $\lambda_j$ and eigenfunctions $\varphi_j(\mathbf{x})$ of $\mathcal{K}_0$. If we denote with $\nu_i$ the eigenvalues of $\mathbf{K}$, the Spectral Mapping Theorem guarantees that the eigenvalues of the Kolmogorov operator can be found as 
\begin{equation}
\label{eq: eigenvalues Kolmo}
    \lambda_k= \frac{\ln \nu_k}{\Delta t}.
\end{equation}
Similarly, the eigenfunctions of $\mathcal{K}_0$ are estimated as 
\begin{equation}
\label{eq: eigenfunction}
    \varphi_k(\mathbf{x}) = \boldsymbol{\Psi}(\mathbf{x})\boldsymbol{\xi}_k,
\end{equation}
where $\boldsymbol{\xi}_k$ are the right eigenvectors of the matrix $\mathbf{K}$. A quite remarkable feature of EDMD, which will be extremely important for the practical implementation of equations \eqref{eq: spectral decomposition corr and Green's functions}, is that it also yields the decomposition of general observables onto the eigenfunctions $\mathcal{\varphi}_k(\mathbf{x})$. 
If we consider any observable $f \in \mathcal{F}_N$, we can write a decomposition onto the dictionary functions as $f(\mathbf{x}) = \sum_{j=1}^N h_j \psi_j(\mathbf{x})=\boldsymbol{\Psi}(\mathbf{x})\mathbf{h}$ where $\mathbf{h}=(h_1,\dots,h_N)^T \in \mathbb{C}^N$ is some vector of constant coefficients.  EDMD then prescribes that the observable $f \in  \mathcal{F}_N$ can be decomposed onto the Kolmogorov eigenfunctions as
\begin{equation}
\label{eq: decomposition of f}
    f(\mathbf{x}) = \sum_k f_k \varphi_k(\mathbf{x}) ,
\end{equation}
where the vector of coefficients $\mathbf{f}$, commonly referred to in deterministic settings as Koopman modes ~\cite{Budisic2012,Brunton2022}, can be found as 
\begin{equation}
\label{eq: from dic to Koopman}
    \mathbf{f} = \mathbf{W}' \mathbf{h},
\end{equation}
where 
\begin{equation}
    \mathbf{W} = [\mathbf{w}_1 \mathbf{w}_2 \dots \mathbf{w}_N] \in \mathbb{C}^{N\times N}
\end{equation}
is the matrix of left eigenvectors $\mathbf{w}_k$ of $\mathbf{K}$ ( i.e. $\mathbf{w}_k^* \mathbf{K} =  \mathbf{w}_k^* \nu_k) $.

\subsection{Kolmogorov spectral reconstruction}
\label{sec: spectral reconstruction}
We here want to leverage the EDMD algorithm to spectrally reconstruct the time-lagged correlations and response functions from data. This approach corresponds to finding a suitable numerical estimate of the mixing rates $\lambda_j$ and projections $\alpha_j(f,g)$ and $\beta_j(f,\Gamma)$ appearing in equations \eqref{eq: spectral decomposition corr and Green's functions} from the EDMD output. 
Since the action of the Markov semigroup on its eigenfunctions is straightforward
\begin{equation}
    e^{t \mathcal{K}_0} \varphi_k(\mathbf{x}) =  e^{t \lambda_k} \varphi_k(\mathbf{x}),
\end{equation}
we seek a decomposition of observables $f = \sum_k f_k \varphi_k$ and $g=\sum_lg_l \varphi_l$ onto the Kolmogorov eigenfunctions. The correlation function \eqref{eq: correlation function} can then be estimated as 
\begin{equation}
\label{eq: spectral decomposition correlation function}
    C_{fg}(t) = \langle g, e^{t\mathcal{K}_0} f \rangle_{\mu_0} = \sum_{k,l}^N f_k g_l^* \langle \varphi_l ,\varphi_k \rangle_{\mu_0}e^{\lambda_k t}=\sum_k \tilde{f}_k e^{\lambda_k t},
\end{equation}
where $\tilde{f}_k=f_k \sum_l g_l^*\langle \varphi_l ,\varphi_k \rangle_{\mu_0}$ and where we have defined the usual $L^2_{\mu_0}$ scalar product as $\langle f, g \rangle_{\mu_0}  = \int f^* (\mathbf{x})g(\mathbf{x})\mathrm{d}\mu_0(\mathbf{x})$. Now, this scalar product can be can be cast in terms of EDMD quantities. Considering equation \eqref{eq: eigenfunction} we can write
\begin{align}
\label{eq: scalar products wrt mu}
    \langle \varphi_l,\varphi_k\rangle_{\mu_0} = \sum^N_{ij} \left( \boldsymbol{\xi}_l^{(i)}\right)^* \boldsymbol{\xi}_k^{(j)}\langle \psi_i , \psi_j\rangle_\mu = \sum^N_{ij} \Xi^*_{il}\Xi_{jk}G_{ij} = \left( \boldsymbol{\Xi}' \mathbf{G} \boldsymbol{\Xi}\right)_{lk},
\end{align}
where we have introduced the matrix of right eigenvectors of $\mathbf{K}$ as 
\begin{equation}
    \boldsymbol{\Xi} = [\boldsymbol{\xi}_1 \boldsymbol{\xi}_2 \dots \boldsymbol{\xi}_N] \in \mathbb{C}^{N\times N}.
\end{equation}
Below, we provide multiple methods to reconstruct the response of the system using EDMD properties. 
\paragraph{Method 1} \mbox{}  \\ 
Since the FDT \eqref{eq: Green Function as correlation} shows that the Green's function is a suitable time-lagged correlation function, a similar result holds for 
\begin{equation}
\label{eq: spectral Green 1 version}
    G_{f}(t) = \Theta(t)\langle \Gamma , e^{t \mathcal{K}_0} f\rangle_{\mu_0} = \Theta(t)\sum_{k,l = 1}^N  \Gamma_l^* f_k  \langle \varphi_l ,\varphi_k \rangle_{\mu_0}e^{\lambda_k t}.
\end{equation}
An important remark has to be made. The previous equation relies on the fact that a reliable estimate of the decomposition of  $\Gamma = \frac{\mathcal{L}_p \rho_0}{\rho_0}$ onto the Kolmogorov eigenfunctions, $\Gamma(\mathbf{x}) = \sum_{k=1}^N \Gamma_k \varphi_k(\mathbf{x})$, is attainable. As explained in the previous section, this entails finding first the decomposition of $\Gamma(\mathbf{x})$ into the dictionary basis and then shifting to the eigenfunctions space through equation \eqref{eq: from dic to Koopman}. \\
We expect this procedure to work well for systems with a smooth enough invariant distribution $\rho_0(\mathbf{x})$, as considered in section \ref{sec: 1D map}.
On the other hand, we expect this strategy not to be reliable when considering weak noise limits of chaotic deterministic systems with a fractal attractor, as considered for example in section \ref{sec: 2D map}. For $\sigma \to 0$, the chaotic system will mostly be concentrated around the underlying chaotic attractor and will not explore the whole phase space unless ultra-long time scales are considered. On a practical level, for a finite dataset $M < \infty$, sampling problems will arise when estimating the invariant distribution $\rho_0(\mathbf{x})$. In these cases, it is not so clear how to numerically regularise the action of the perturbation operator $\mathcal{L}_p$ and, more importantly, the division by $\rho_0$ when estimating the observable $\Gamma(\mathbf{x})$. We anticipate that the estimation of the Green's function of the second system investigated in section \ref{sec: 2D map} using this procedure heavily depends on the numerical regularisation choices. 
We thus propose two alternative reconstruction strategies that do not require similar regularisation procedures.
\paragraph{Method 2} \mbox{} \\
The Green's function \eqref{eq: Green Function as correlation} can be rewritten as a scalar product with respect to the standard Lebesgue measure 
\begin{equation}
\label{eq: Green's function alternative}
    G(t) = \Theta(t) \langle \mathcal{L}_p \rho_0 , e^{\mathcal{K}_0t} f  \rangle \coloneqq \Theta(t) \langle \gamma , e^{\mathcal{K}_0t} f  \rangle,
\end{equation}
where in the last equality we have defined the observable $\gamma = \mathcal{L}_p \rho_0$ and $\langle f, g \rangle =  \int_{\mathcal{M}} f^*(\mathbf{x})g(\mathbf{x})\mathrm{d}\mathbf{x} $ is the standard scalar product. On the one hand, the previous expression provides a way to evaluate $G_f(t)$ starting from the estimation of the much smoother observable $\gamma$ rather than $\Gamma$. On the other hand, it requires one to perform an integral uniformly over the whole phase space $\mathcal{M}$. Since EDMD approximates the properties of the operator $\mathcal{K}$ in terms of globally defined dictionary basis functions, the problem of evaluating \eqref{eq: Green's function alternative} can be mapped into the much simpler problem of computing standard scalar products among the dictionary functions. An analogous calculation that leads to equation \eqref{eq: spectral Green 1 version} shows that the Green's function can also be decomposed as
\begin{equation}
\label{eq: spectral Green 2 version}
    G_f(t) = \Theta(t) \sum_{k,l}^Ne^{\lambda_k t} f_k \gamma_l^* \langle \varphi_l,\varphi_k\rangle.
\end{equation}
The standard scalar product among the Kolmogorov eigenfunctions can be written in terms of EDMD-related quantities as
\begin{equation}
\label{eq: scalar product with respect to Lebesgue}
    \langle \varphi_l, \varphi_k \rangle = \left( \boldsymbol{\Xi}' \mathbf{G}^{Leb} \boldsymbol{\Xi}\right)_{lk},
    \end{equation}
where 
\begin{equation}
    \mathbf{G}^{Leb}_{ij} = \langle \psi_i , \psi_j \rangle 
\end{equation}
is the matrix of scalar products of dictionary functions. In particular, if an orthonormal basis of dictionary functions is chosen, then $\mathbf{G}^{Leb}_{ij} = \delta_{ij}$.
\paragraph{Method 3} \mbox{} \\
Both methods 1 and 2 require a binning procedure to evaluate an empirical estimate of the invariant density $\rho_0$ in order to reconstruct the action of the perturbation operator $\mathcal{L}_p \rho_0$, which requires a large amount of data. However, it is possible to obtain projections of the response observable $\Gamma$ on the Kolmogorov eigenfunctions directly from the time series data originating from the dynamical system. In method 1 we seek a global approximation of the observable $\Gamma$ in terms of the dictionary functions $\psi_i(\mathbf{x})$. Here, we propose to find an approximation of $\Gamma$ in terms of the dictionary by weighting areas of phase space according to the invariant measure. Mathematically, we seek an approximation $\Gamma_{approx}(\mathbf{x})=\sum_i c_i \psi_i(\mathbf{x})$ such that the quadratic objective function $|\Gamma(\mathbf{x}) - \Gamma_{approx}(\mathbf{x})|_\nu^2 = \langle \Gamma - \Gamma_{approx}, \Gamma - \Gamma_{approx} \rangle\nu$ is minimised. In the previous expression, $\nu$ is a suitable measure. In Method 1 we looked for a global approximation of $\Gamma$ by setting $\nu$ as the Lebesgue measure. We here choose $\nu=\mu_0$, so that areas where the system spends most time are weighted more than areas that are rarely visited. Minimising the quadratic function with $\nu=\mu_0$ results in the following parameter vector 
\begin{equation}
\label{eq: coefficient vector method 3}
    \mathbf{c} = \mathbf{G}^+ \boldsymbol{\Delta}
\end{equation}
where $\mathbf{G}_{ij} = \langle \psi_i , \psi_j \rangle_{\mu_0}$ is the Gram matrix of scalar products between dictionary functions which is estimated through the EDMD procedure, and $\boldsymbol{\Delta}_i = \langle \psi_i , \Gamma \rangle_{\mu_0}$. Now, we estimate this scalar product as 
\begin{equation}
\label{eq: method 3 projections}
    \boldsymbol{\Delta}_i =  - \int_{\mathcal{M}} \nabla \cdot \left( \mathbf{X}(\mathbf{x}) \rho_0 \right)\psi^*_i(\mathbf{x})\mathrm{d}\mathbf{x} = \int_{\mathcal{M}} \mathbf{X}(\mathbf{x})\cdot \nabla \psi^*_i \rho_0(\mathbf{x}) \mathrm{d}\mathbf{x} \approx \frac{1}{T} \int_0^T \mathbf{X}(\mathbf{x}(t))\cdot \nabla \psi^*_i(\mathbf{x}(t))\mathrm{d}t
\end{equation}
where in the last equality we have used ergodicity to approximate the expectation value as a time average on a long ($T \to +\infty$) trajectory. The previous expression, together with \eqref{eq: coefficient vector method 3}, provide a simple way to estimate the decomposition of $\Gamma$ onto the dictionary functions from time averages of suitable observables involving the derivative of the dictionary functions. We observe that, if the dictionary is closed under differentiation, the derivative $\nabla \psi_i$ can be exactly represented in terms of the dictionary itself and thus Method 3 does not require any additional information than the EDMD procedure itself. 
\paragraph{Summary of the Kolmogorov Spectral Reconstruction procedure} \mbox{} \\
We provide here, for convenience, a summary of the key steps of the spectral reconstruction procedure, leaving the specific details to section \ref{sec: numerical experiments}. \begin{itemize}
    \item Consider an evenly sampled trajectory of the stochastic system \eqref{eq: Stochastic Differential equations} as $\{ \mathbf{x}_i = \mathbf{x}(t = i \Delta t) \}_{i=1}^M$.
    \item Select a dictionary $\boldsymbol{\Psi}(\mathbf{x})$ for EDMD and feed the algorithm with the snapshot data $\mathbb{D} = \{(\mathbf{x}_i,\mathbf{y}_i = \mathbf{x}_{i+1}) \}$. Construct matrices $\mathbf{G}$, $\mathbf{A}$ and the Kolmogorov matrix $\mathbf{K}$. 
    \item The output of EDMD is the triple $\left(\{\lambda_k\}_{k=1}^N,\boldsymbol{\Xi}, \mathbf{W}\right)$ of eigenvalues, right eigenvector matrix and left eigenvector matrix. 
    \item Given any observable $f \in \mathcal{F}_N$, find its decomposition onto the dictionary as $f(\mathbf{x}) = \boldsymbol{\Psi}(\mathbf{x})\mathbf{h}$. The coefficients $\mathbf{f}$ of the decomposition of $f(\mathbf{x})$ onto the Kolmogorov eigenfunctions are then obtained as $\mathbf{f}= \mathbf{W}'\mathbf{h}$. 
    \item Spectrally reconstruct the time-lagged correlation function $C_{fg}(t)$ of any two observables $f$ and $g$ with equations \eqref{eq: spectral decomposition correlation function} and \eqref{eq: scalar products wrt mu}.
    \item Method 1 and 2: for the evaluation of Green's functions,  approximate the invariant density $\rho_0$ from the time series $\{ \mathbf{x}_i \}_{i=1}^M$. Evaluate the action of the perturbation operator $\mathcal{L}_p$ onto $\rho_0$ and construct observable $\Gamma$ (or $\gamma$). Method 3: Evaluate the moments $\boldsymbol{\Delta}$ from time averages with \eqref{eq: method 3 projections}. 
    \item The Green's function can then be estimated with equation \eqref{eq: spectral Green 1 version} for Method 1 and 3 (or \eqref{eq: spectral Green 2 version} for Method 2).
\end{itemize}

\section{Numerical experiments}
\label{sec: numerical experiments}
As the first two examples, we investigate the mixing and response properties of two chaotic maps. Even though the theoretical results presented above apply to continuous time stochastic dynamical systems, the situation for maps is similar, if not easier, as there is no time-sampling of trajectories associated with it. In fact, all the spectral decomposition formulas hold when the eigenvalues $\lambda_k$, see equation \eqref{eq: eigenvalues Kolmo}, are evaluated with $\Delta t = 1$. 

In each of the cases investigated below, we take the luxury of observing the full phase space of the system.   In all the below examples, the EDMD-reconstructed statistical properties have been compared to numerical experiments obtained on a different dataset than the one employed for EDMD. Code for the numerical experiments can be found at \url{https://github.com/niccozagli/KoopmanismResponse}.

\subsection{One-dimensional uniformly expanding map}
\label{sec: 1D map}
\begin{figure}
     \centering
     \begin{subfigure}[b]{0.49\textwidth}
         \centering
         \includegraphics[width=\textwidth]{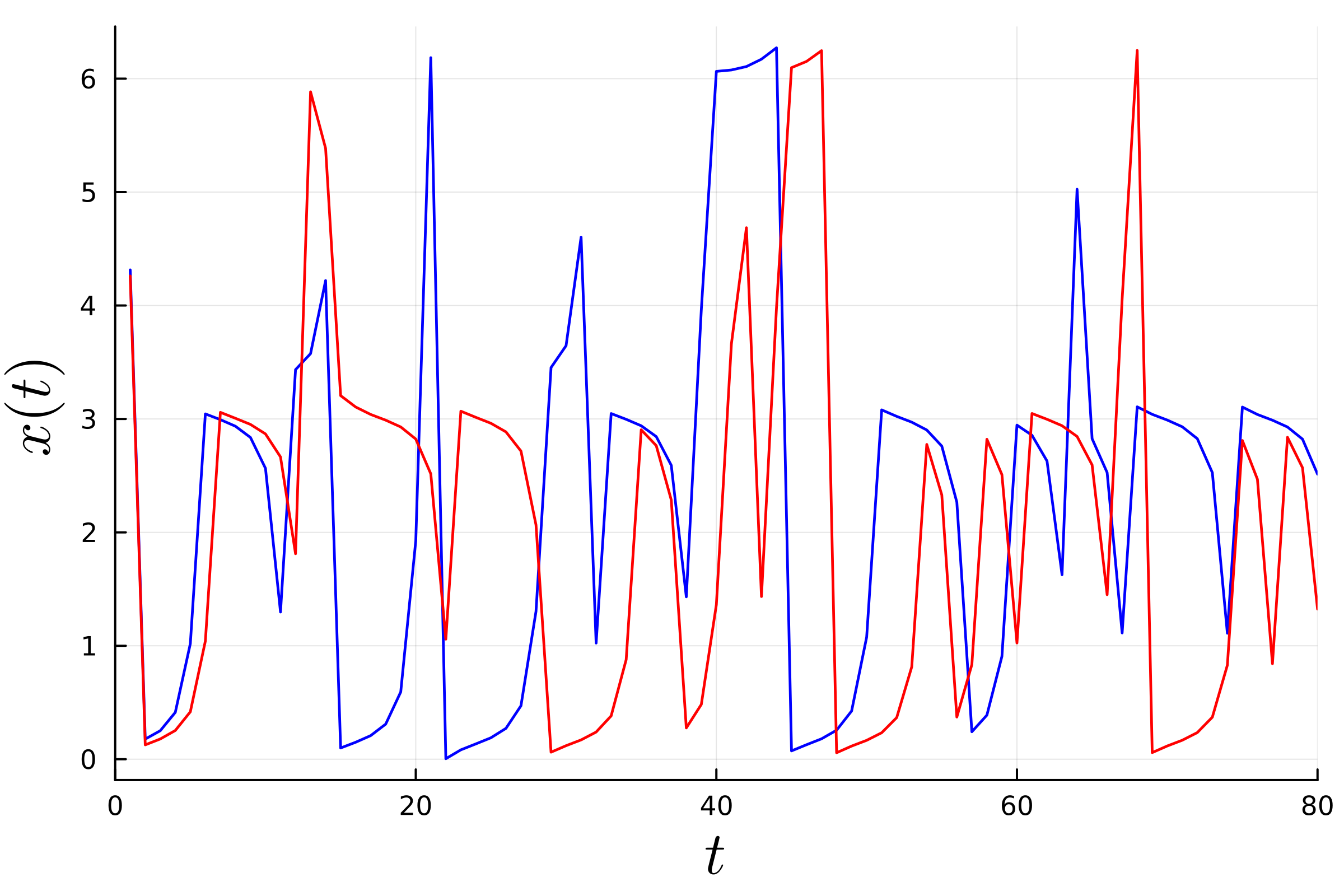}
               \caption{}
     \end{subfigure}
     \hfill
     \begin{subfigure}[b]{0.49\textwidth}
         \centering
         \includegraphics[width=\textwidth]{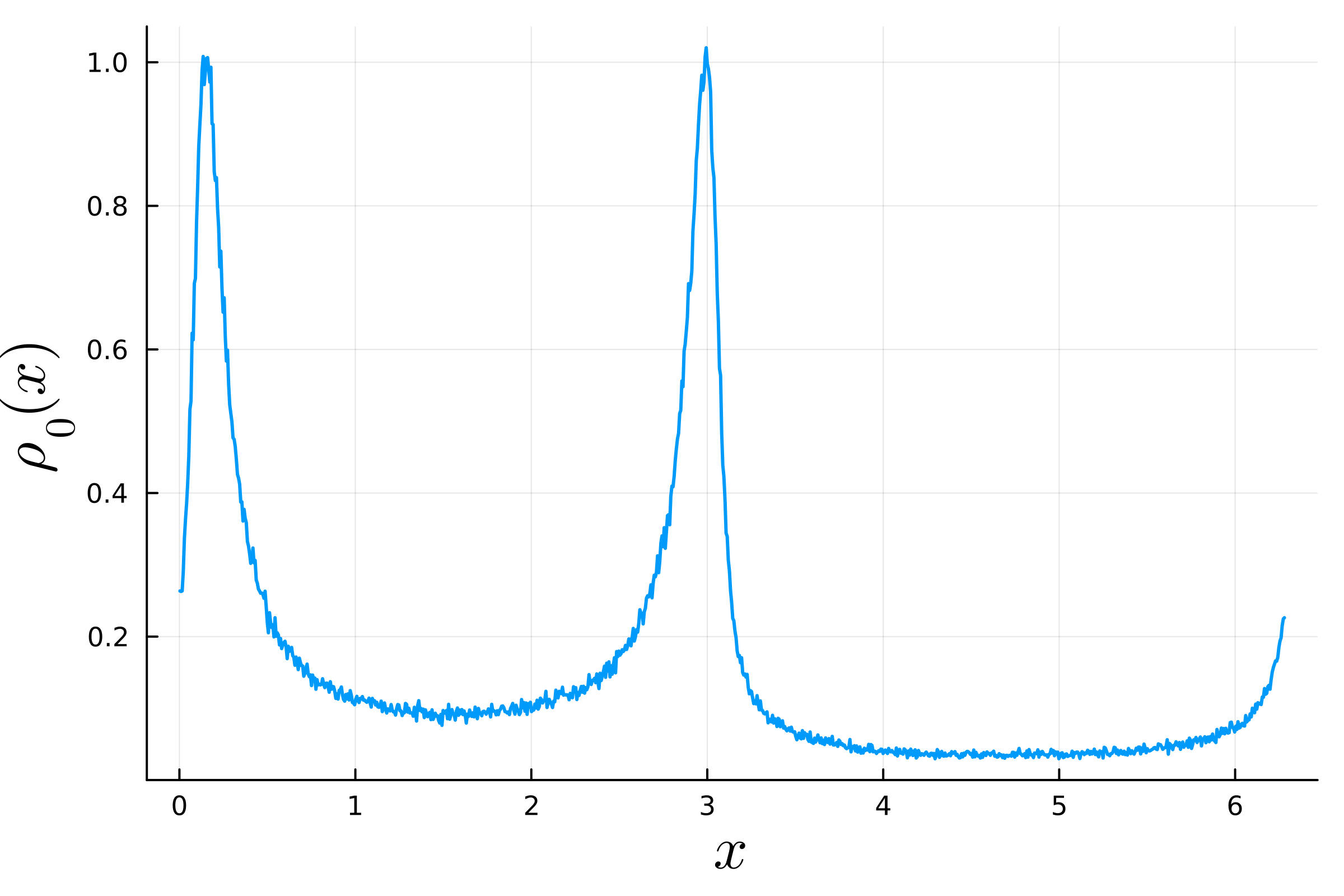}
         \caption{}
         \label{fig: 1}
     \end{subfigure}
    
        \caption{Panel (a): Chaotic Trajectories. Panel (b): Invariant Density computed from a long chaotic trajectory of the system. The density will be used to compute the response properties of the system, see later discussion.}
        \label{fig: First}
\end{figure}

We will first consider the simplest settings where chaotic dynamics can arise, that is one-dimensional uniformly expanding maps. Despite their simple structure, such maps represent the first examples for the analytical investigation of chaotic phenomena \cite{B00}. We consider the following map supported on the torus $\mathcal{M}=\mathbb{T}_{[0,2\pi)}=\mathbb{R}/(2\pi\mathbb{Z})$, introduced in ~\cite{Wormell2025_EDMD}
\begin{equation}
\label{eq: 1d map}
    x_{n+1} = F(x_n) \coloneqq \alpha x_n - \gamma \sin\left(6x_n \right)+\Delta \cos\left(3x_n \right) \mod 2 \pi ,
\end{equation}
with $\alpha = 3$, $\gamma = 0.4$ and $\Delta = 0.08$. Two typical chaotic trajectories originating from nearby initial conditions are shown in panel (a) of Figure \ref{fig: First}.
At variance with the theory developed in the previous section, this dynamical system is completely deterministic. In these settings, the spectral decomposition of the Markov semigroup in $L^2_{\mu_0}$ is no longer valid, see the discussion following Eq. \ref{eq: spectral decomposition semigroup}. 
Nevertheless, we propose this example as it highlights the importance of the smoothness properties of the invariant measure $\mu_0$.  Being a uniformly expanding and analytic map, its invariant measure is guaranteed to be absolutely continuous with respect to Lebesgue with smooth density $\rho_0(\mathbf{x})$, see also Panel (b) of Figure \ref{fig: First}. On the one hand, this supports the validity of the FDT \eqref{eq: Green Function as correlation}. On the other hand, for this class of maps having an analytic measure, there is numerical and analytical evidence that EDMD, with a specific choice of dictionary functions, provides eigenvalues which converge in the limit of infinite dictionary to the (deterministic) Ruelle-Pollicott resonances ~\cite{Wormell2025_EDMD}, thus expressing the decay of correlation functions of the chaotic dynamics. Note that such resonances do not correspond to the eigenvalues of the Koopman operator in $L^2(\mu_0)$, which are instead located in the unit circle. The non-trivial link between the spectrum of the Koopman operator in non-standard function spaces and the Ruelle-Pollicott resonances for deterministic hyperbolic systems on the torus has been investigated in \cite{ANTONIOU1997,Slipantschuk_2017,Pollicott_2023}.

\begin{figure}
    \centering
    \begin{subfigure}[b]{0.49\textwidth}
         \centering
    \includegraphics[width=\textwidth]{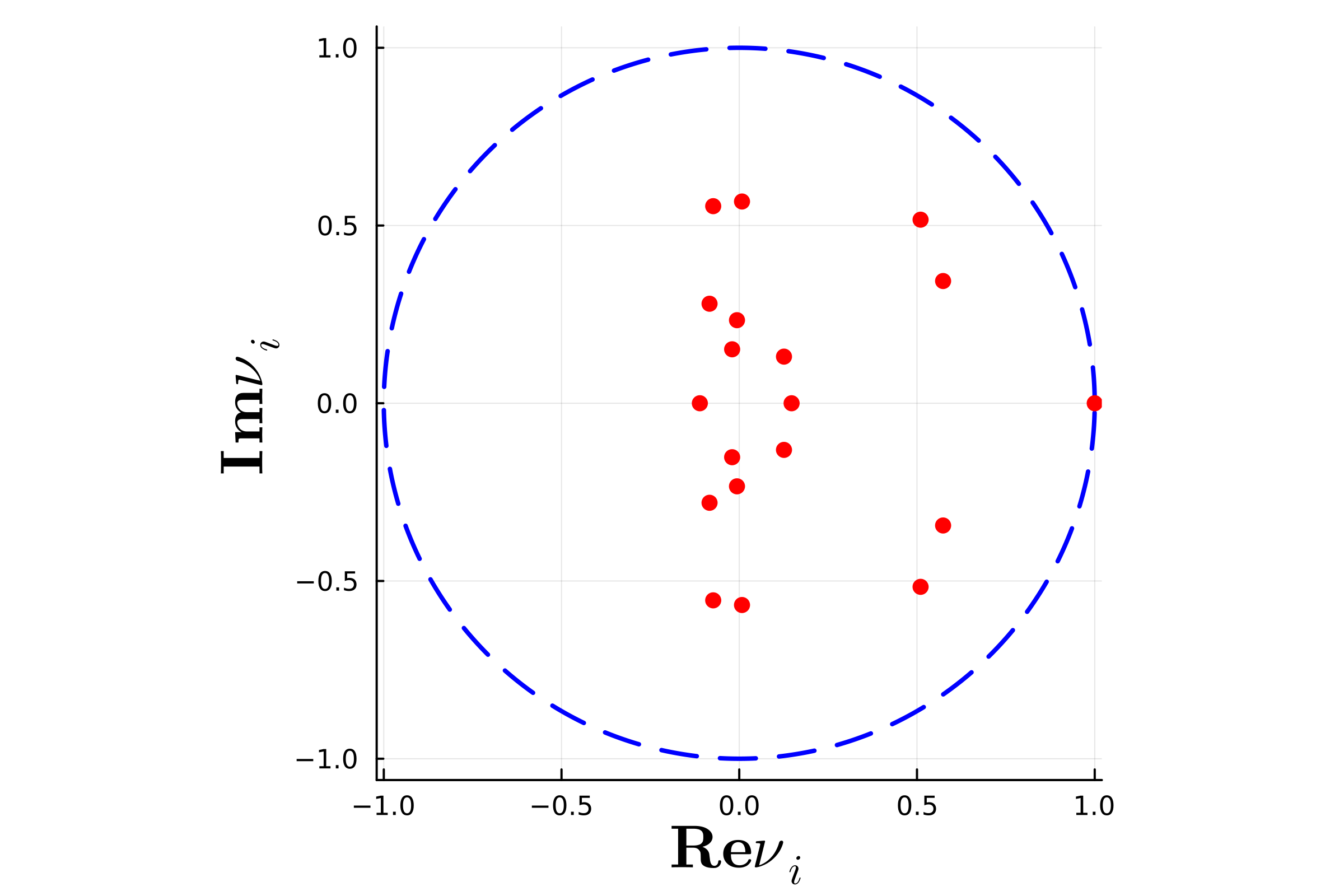}
         \caption{}
         \label{fig: 2a}
     \end{subfigure}
     \hfill
      \begin{subfigure}[b]{0.49\textwidth}
         \centering
    \includegraphics[width=\textwidth]{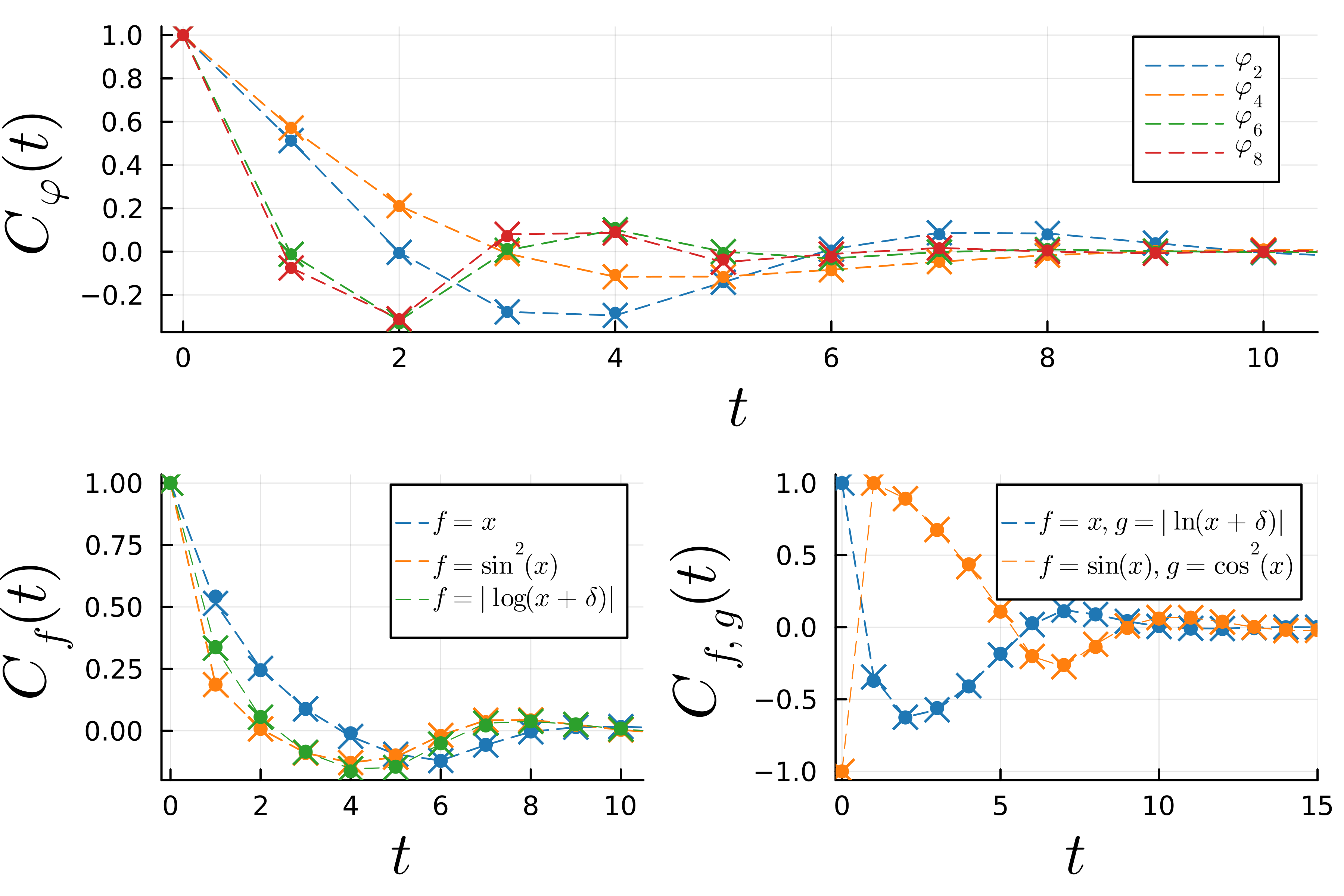}
         \caption{}
         \label{fig: 2b}
     \end{subfigure}
    \caption{Panel (a):  Deterministic Ruelle-Pollicott resonances estimated using EDMD with a Fourier basis. Only the first 20 resonances are shown. Panel (b): Correlation Functions of various observables. Top: correlation functions of $\mathcal{K}_0-$eigenfunctions $\varphi_k(x)$. The crosses correspond to the numerical simulations, the dots to the corresponding exponential term $|\nu_i|^t$. Bottom: correlation functions of general observables. Crosses correspond to numerical simulations, the dots to the spectral decomposition. All correlation functions have been normalised by their initial value for scale. The dashed lines are included only for visualisation purposes. Here $M=5\times 10^5$ and $K_{max}=30$. }
    \label{fig:Correlation Functions}
\end{figure}

\begin{figure}
     \centering
     \begin{subfigure}[b]{0.49\textwidth}
         \centering
         \includegraphics[width=\textwidth]{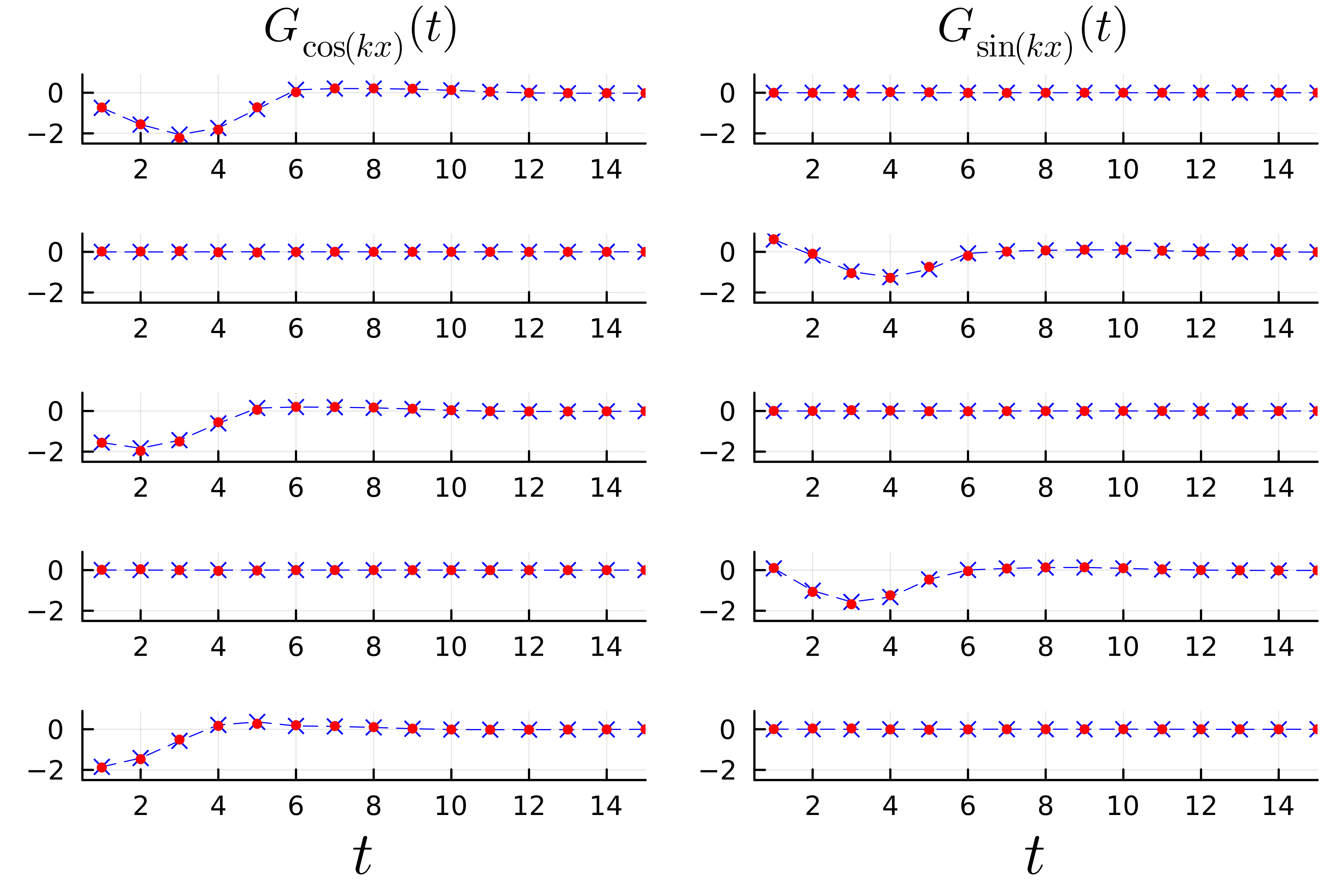}
         \caption{}
         \label{fig: Third 1}
     \end{subfigure}
\hfill
\begin{subfigure}[b]{0.49\textwidth}
         \centering
         \includegraphics[width=\textwidth]{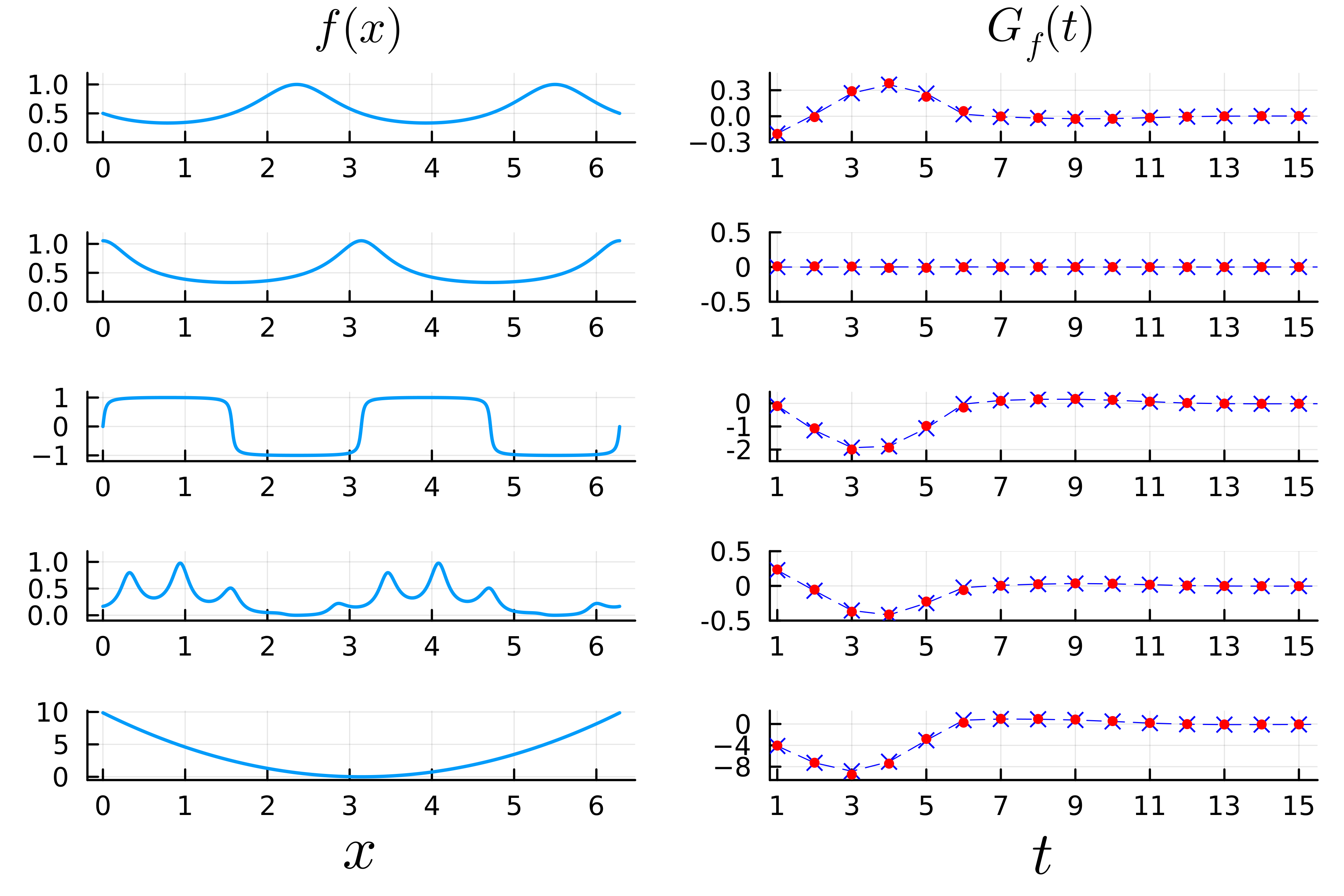}
         \caption{}
         \label{fig: Third 2}
     \end{subfigure}
        \caption{Green's functions corresponding to a uniform perturbation $X(x) =1$. Panel (a): Green's functions associated with the first plane waves $(\sin(kx),\cos(kx))$ with $k=1,\dots,5$ (from top to bottom). Panel (b): Left column shows different observables with their corresponding Green's functions on the right column. Blue crosses correspond to direct numerical simulations and red dots to the spectral reconstruction. Here, $M$ and $K_{max}$ are as in Figure \ref{fig:Correlation Functions}.}
        \label{fig: Third}
\end{figure}

These considerations point to the fact that the EDMD-related spectral decompositions \eqref{eq: spectral decomposition correlation function} and \eqref{eq: spectral Green 1 version} might be successful in evaluating the mixing and response properties of the system. The a-posteriori validity of our results, shown in Figs. \ref{fig:Correlation Functions} and \ref{fig: Third}, corroborates this assumption. 
Following ~\cite{Wormell2025_EDMD}, we consider a dictionary of orthonormal Fourier functions, which is a quite natural choice for a system defined on  a periodic domain,
\begin{equation}
    \boldsymbol{\Psi}(\mathbf{x}) = \big( e^{-i K_{max}x}, \dots, 1 , \dots , e^{i K_{max}x}\big)
\end{equation}
The corresponding eigenvalues $\nu_i$ of the EDMD matrix $\mathbf{K}$ are shown in Panel (a) of Figure \ref{fig:Correlation Functions}. Since we consider a discrete time system, the decay rate of the Kolmogorov modes is given by $|\nu_i|$ whereas their oscillatory behaviour is determined by $\arg(\nu_i)$.  EDMD returns the static eigenvalue $\nu_0 =1$, which is a positive self-consistency check as we included its associated constant eigenfunction $\varphi_0 = 1$ in the dictionary. If the interpretation at the basis of the spectral decomposition \eqref{eq: spectral decomposition correlation function} is correct, the time-lagged correlation functions associated with the eigenfunctions of  $\mathcal{K}_0$ are given by a pure exponential decay
\begin{equation}
\label{eq: pure exp decay}
    C_{\varphi_k}(t) = c |\nu_k|^t  = c e^{\lambda_k t} ,
\end{equation}
with $c$ being a constant. We have verified this by estimating the eigenfunctions with EDMD as $\varphi_k(x) = \boldsymbol{\Psi}(x) \boldsymbol{\xi}_k$ and evaluating their correlation functions along a long chaotic trajectory. Panel (b) (top plot) of Figure  \ref{fig:Correlation Functions} shows excellent agreement between the predicted theoretical exponential decay \eqref{eq: pure exp decay} and the numerical results obtained with EDMD. 
We proceed to evaluate auto-(cross-)correlation functions of general observables $f$ and $g$ of the system, shown in the bottom left (right) plot in Panel (b) of Figure \ref{fig:Correlation Functions}. Most of these observables, not necessarily having a physical meaning, do not lie in the space $\mathcal{F}_N$ spanned by the dictionary $\boldsymbol{\Psi}(x)$. We have actually tested the methodology on observables that are not necessarily periodic on $\mathbb{T}_{[0,2\pi)}$. We have only required that they do not exhibit any divergences, for example, we have employed a regularisation $f(x) = |\log(x+\delta)|$ with $\delta = 0.1$. 
\\\\
Having confirmed that the mixing properties of the system can be numerically reconstructed through the EDMD spectral reconstruction methodology, we have investigated the more complicated and subtle issue of reconstructing the response properties of the system. Perturbing equation \eqref{eq: 1d map} consists of setting $F(x_n) \to F(x_n) + \varepsilon X(x_n) T(t)$. Here, in particular, we consider the effect of a uniform perturbation $X(x) \equiv 1$ $\forall x$. As observed before, its associated response-related observable $\Gamma(x) =  \frac{\mathcal{L}_p \rho_0}{\rho_0} =  - \frac{\partial }{\partial x}\ln \rho_0$ is intrinsic to the problem and is not a prior known as there is no analytic expression for the invariant measure of the system. We estimate from a chaotic trajectory the invariant density of the unperturbed system $\rho_0(x)$, see Panel (b) of Figure \ref{fig: First}, and construct the observable $\eta(x) = \ln \rho_0$. Since the Fourier transform commutes with the derivative operator, we can interchangeably take the derivative of $\eta(x)$ to obtain $\Gamma(x)$ and then project onto the dictionary $\boldsymbol{\Psi}(x)$ or vice-versa. We proceed with the latter option as it allows us to take a spectral derivative as follows. We first look for a decomposition onto the dictionary $\eta(x) = \sum_k \eta_k \psi_k(x)$, take the derivative as $\Gamma(x) = \sum_k i k \eta_k \psi_{k}(x) \coloneqq \sum_k h_k \psi_k(x)$ and then find its associated decomposition onto the $\mathcal{K}_0-$eigenfunctions as $\Gamma(x) = \sum_k \Gamma_k \varphi_k(x)$ with $\Gamma_k = (\mathbf{W}\mathbf{h})_k$ as prescribed by equation \eqref{eq: from dic to Koopman}. \\
Having estimated the decomposition of $\Gamma$ onto the $\mathcal{K}_0-$eigenfunctions, we now reconstruct the Green's function of various observables of the system. Linear response theory applies only to continuous observables, so we here test our methodology on smooth periodic functions on $\mathbb{T}_{[0,2\pi)]}$. We compare the spectral reconstruction of the Green's function with direct numerical results averaged over a large number ($5\times 10^6$) of response experiments. Figure \ref{fig: Third} exhibits an excellent agreement between the spectrally reconstructed Green's function and its numerically evaluated counterpart. In panel (a) we provide the Green's functions associated to the basis of periodic functions on $\mathbb{T}_{[0,2\pi)}$. We observe a quite peculiar symmetry where the observables $\cos(kx)$ ($\sin(kx)$) $k=1,\dots,5$ have vanishing linear response for $k$ even (odd). We have also tested the spectral reconstruction on observables $f \notin \mathcal{F}_N$ that do not belong to the span of the dictionary. In particular, panel (b) shows the response for the following observables $f(x) = (2+\sin(2x))^{-1}$,$ f(x)=c \cos\left( \arctan\left( 3 \sin(x)\right)\right)$, $f(x)=c \arctan(20\sin(2x))$, $f(x)=\frac{1}{2}(1+\sin(2x))(2+\cos(10x))^{-1}$, $f(x)= (x-\pi)^2$, where $c$ represents appropriate constants. Even though these observables do not belong to $\mathcal{F}_N$, there is very good agreement with the direct numerical experiments. 
\\
Next, we test the robustness of the spectral reconstruction methodology as the hyperparameters of the problem, such as the length of the trajectory $M$ and the number of dictionary functions $K_{max}$, vary. We show here the results for the observable $f(x) = (x- \pi)^2 \notin \mathcal{F}_N$. We first fix a long trajectory, $M=6\times10^5$, and study the performance of the spectral reconstruction of $G_f(t)$ by varying $K_{max}$.  
\begin{figure}
    \centering
    \begin{subfigure}[b]{0.49\textwidth}
    \centering
    \includegraphics[width=\textwidth]{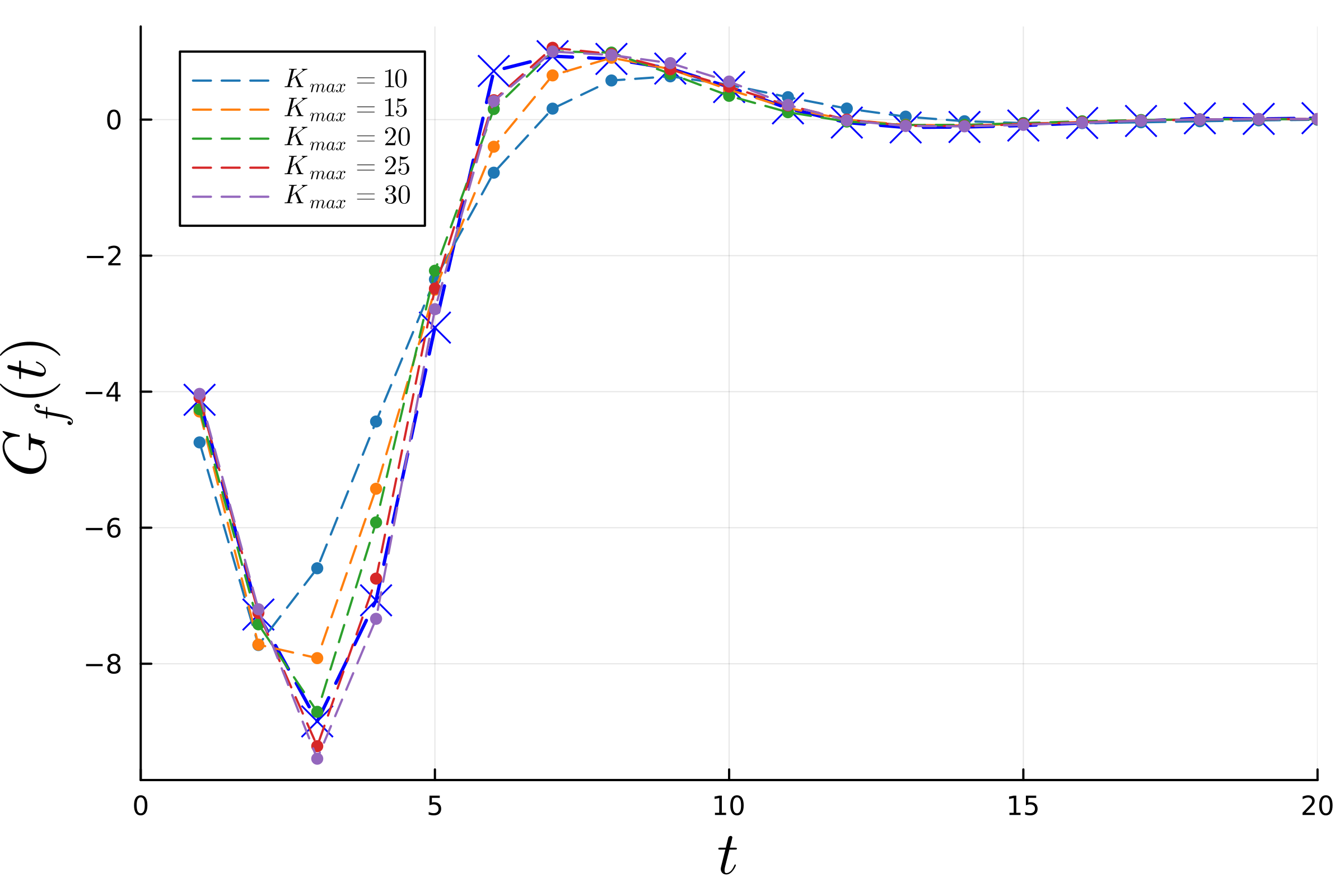}
        \caption{}
    \end{subfigure}
    \hfill
    \begin{subfigure}[b]{0.49\textwidth}
    \centering
    \includegraphics[width=\textwidth]{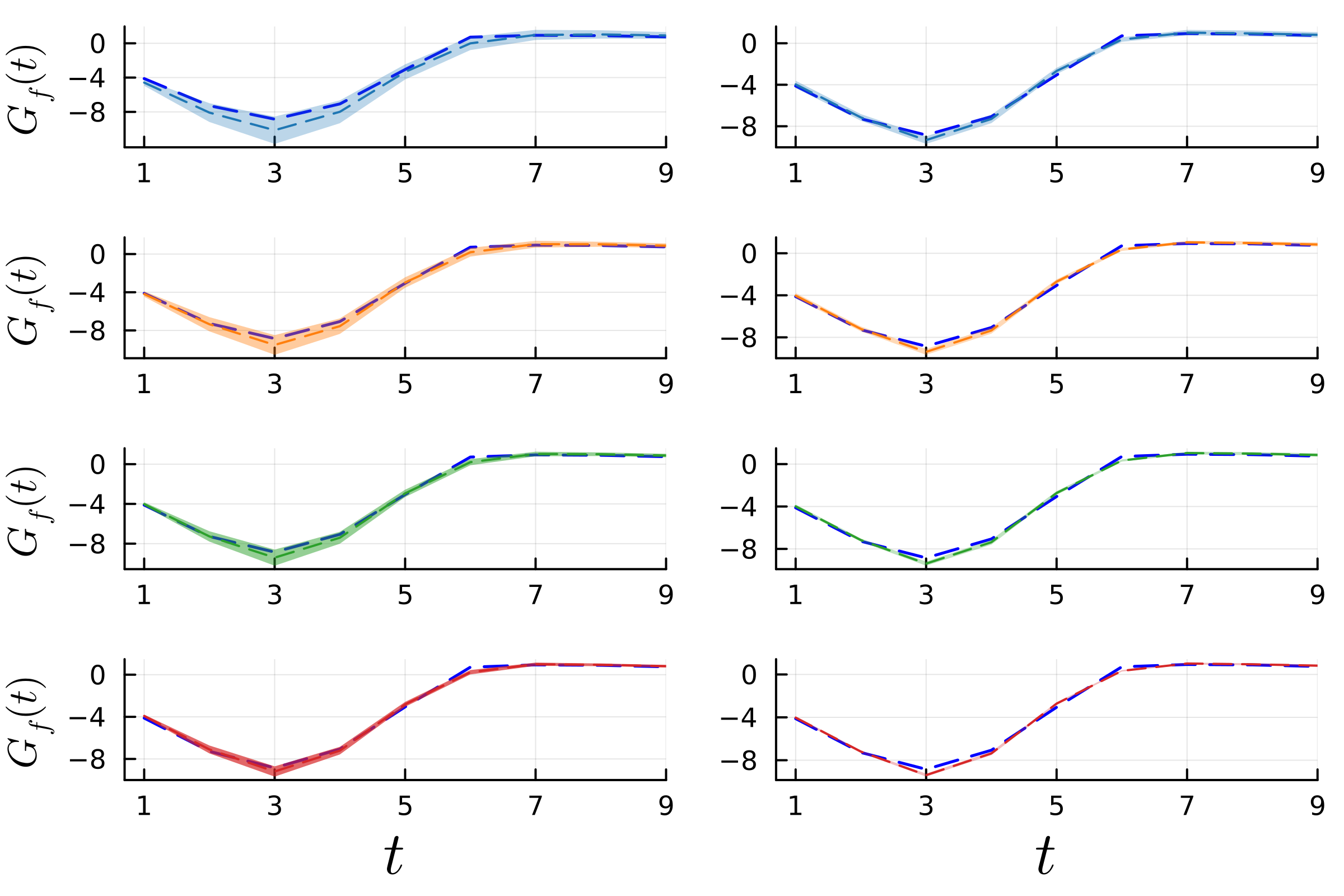}
        \caption{}
    \end{subfigure}
    \caption{Robustness of the spectral reconstruction of response properties. Panel (a): robustness to variations in $K_{max}$. Graphic conventions are as in Figure \ref{fig: Third}. Different colours refer to different $K_{max}$. Panel(b): robustness to variations in $M$. For visualisation purposes, we have dropped the marker (dots and crosses).  The shaded area represents the error bars at each point in time. From top to bottom, $M = 2,4,8,16 \times 10^3$.  Left column: spectral properties of $\mathcal{K}_0$ and $\rho_0$ are both evaluated on $M$ data points. Right column: here $\rho_0$ is evaluated on a long trajectory of length $M_{long} = 6  \times 10^5$ points.  }
    \label{fig:Robustness}
\end{figure}
Panel (a) of Figure \ref{fig:Robustness} shows that, as expected, having a richer dictionary results in a more accurate spectral reconstruction of the Green's function associated with an observable $f \notin \mathcal{F}_N$. A more insightful understanding is gained when considering the scenario where, fixed a dictionary of $K_{max}=30$ modes, the spectral decomposition is performed on trajectories of different lengths $M$. In particular, fixed $M$, we assess the statistical properties of the spectral reconstruction methodology by performing an ensemble of $100$ different experiments corresponding to trajectories stemming from different initial conditions on the attractor. For each trajectory, we evaluate an estimate of the Green's function $G_f(t)$ obtained through the spectral decomposition. This procedure not only gives an estimate of the Green's function, obtained as the mean over the different ensemble members but also provides a quantitative estimate of the statistical uncertainty related to having a short trajectory. Panel (b) (left column) of Figure \ref{fig:Robustness} shows that this estimator performs well on average (dashed lines) even on short trajectories but such trajectories are associated with a bigger uncertainty (shaded area). The origin of this uncertainty is twofold. On the one side, the EDMD algorithm, being a Galerkin method, is expected to poorly perform on shorter trajectories.  On the other side, such uncertainty could arise from a poor representation of the invariant density $\rho_0(x)$ which is a key quantity in the determination of response properties. 
\\
We provide numerical evidence that the biggest part of the uncertainty is given by the latter feature, see the right column of Panel (b). This plot shows the results of the same numerical procedure described above, with the caveat that the unperturbed invariant density $\rho_0(x)$ is now being estimated on a single long trajectory of length $M_{long} = 6\times10^5$. This shows that the properties of $\mathcal{K}_0$ are accurately captured with EDMD from a relatively short trajectory and that the uncertainty regarding response properties is greatly reduced if a separate approximation of the unperturbed measure from a longer trajectory is available.
 \subsection{Generalised Arnold's Cat Map}
\label{sec: 2D map}
\begin{figure}
     \centering
     \begin{subfigure}[b]{0.49\textwidth}
         \centering
\includegraphics[width=\textwidth]{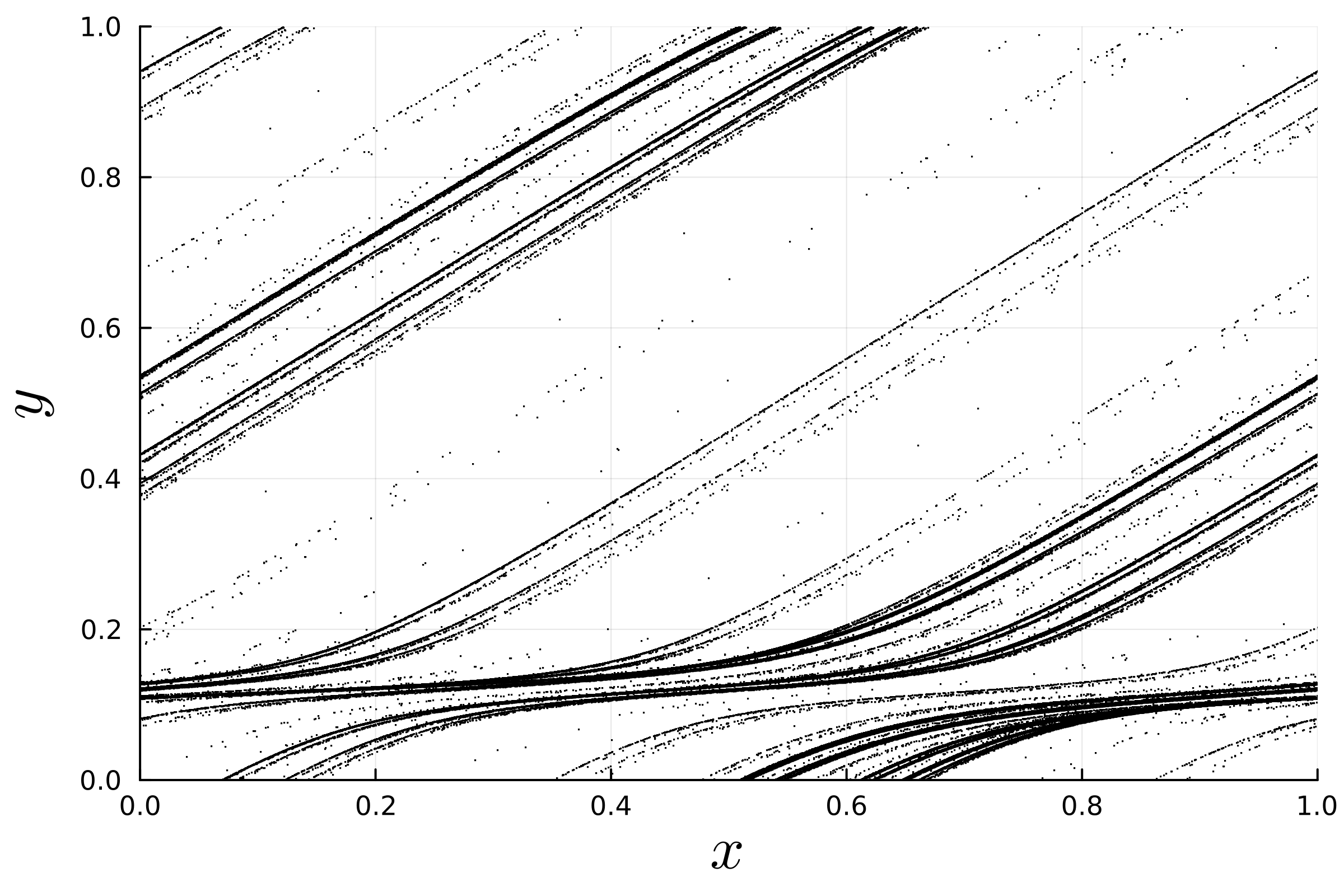}
         \caption{}
         \label{fig: Inv Measure 2D 1}
     \end{subfigure}
     \hfill
     \begin{subfigure}[b]{0.49\textwidth}
         \centering
         \includegraphics[width=\textwidth]{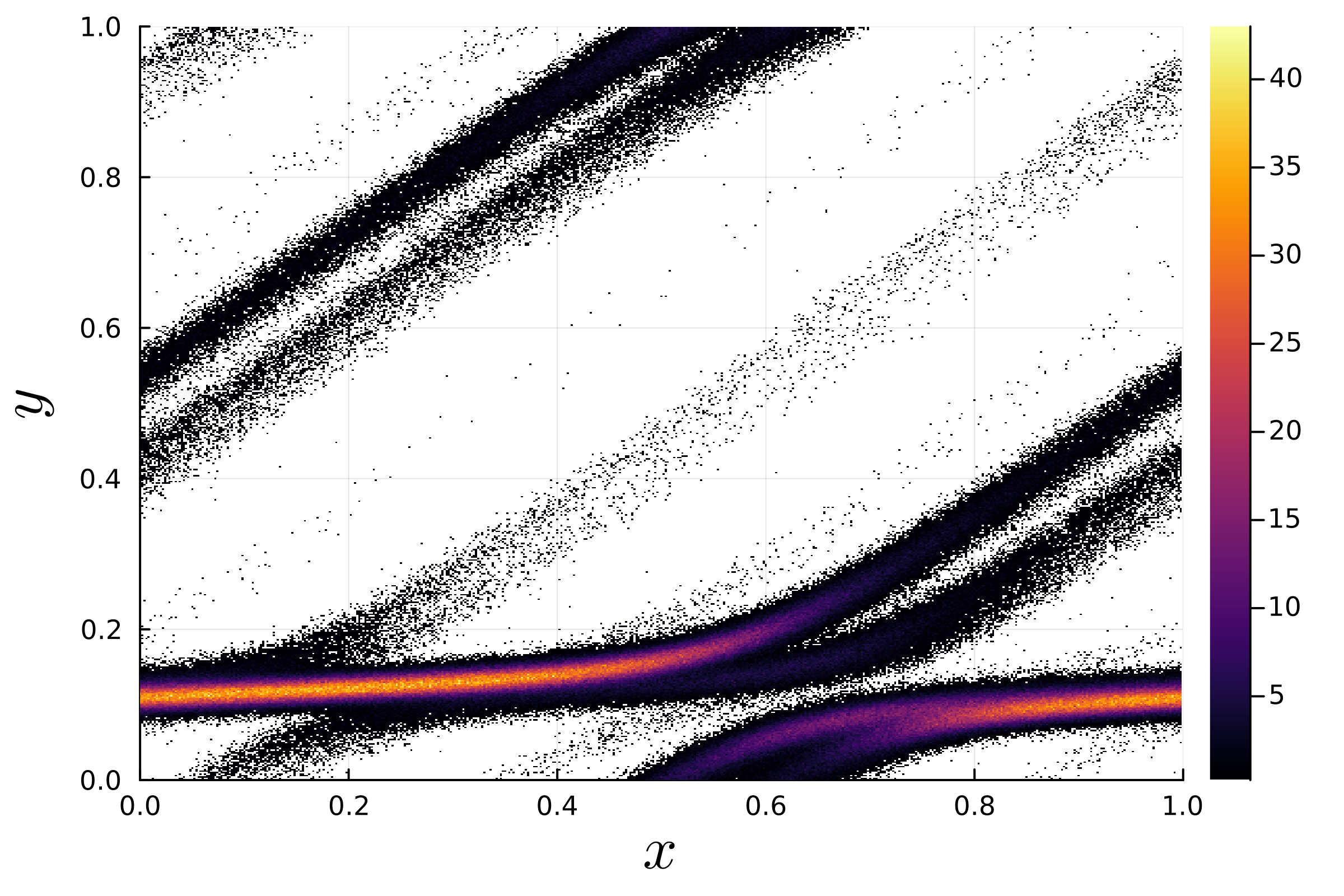}
         \caption{}
         \label{fig: Inv Measure 2D 2}
     \end{subfigure}
        \caption{Panel (a): Deterministic fractal attractor for the generalised Arnold map. Panel (b): Invariant Density $\rho_0(\mathbf{x})$}
        \label{fig: Invariant Measure 2d Map}
\end{figure}

\begin{figure}
     \centering
     \begin{subfigure}[b]{0.49\textwidth}
         \centering
\includegraphics[width=\textwidth]{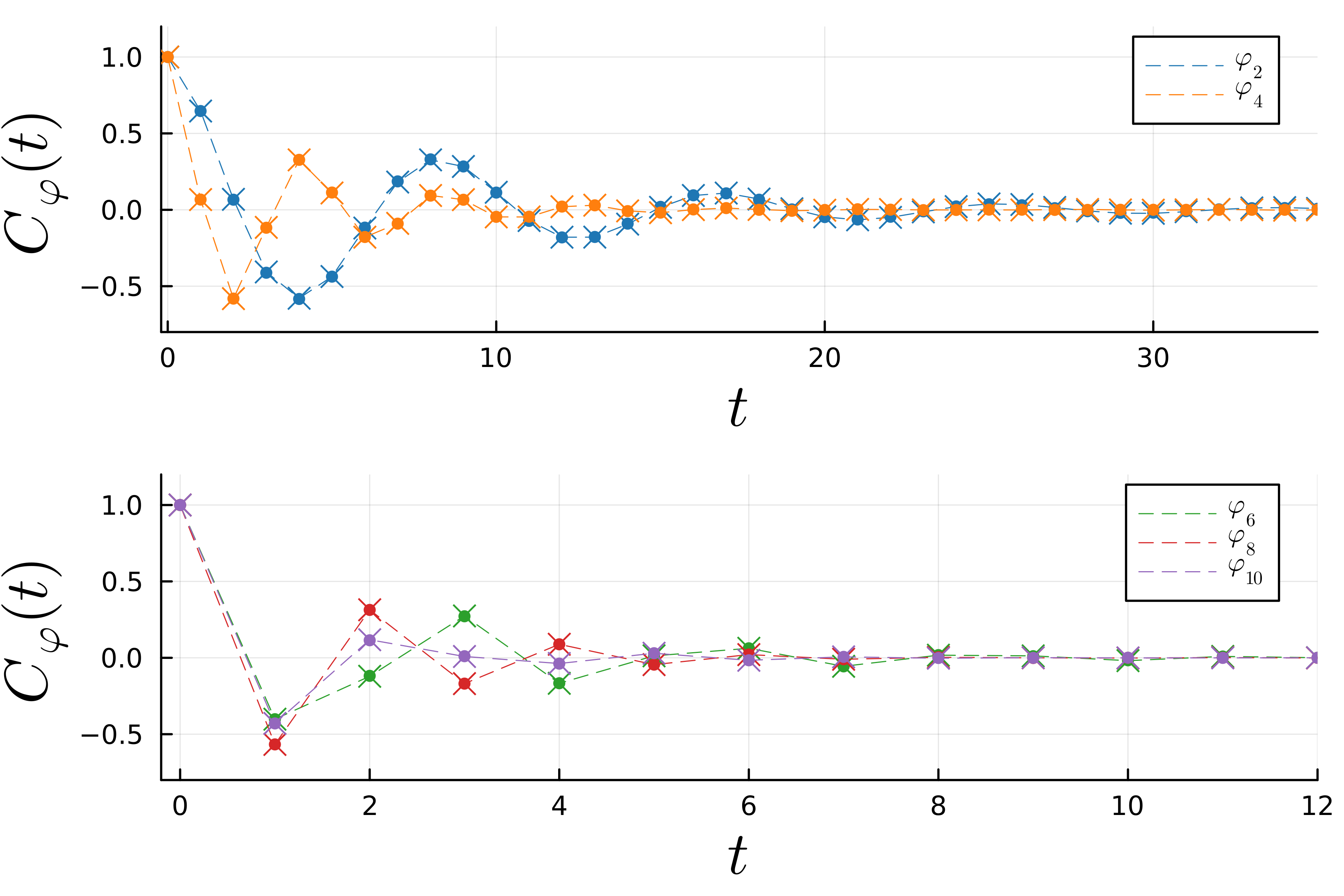}
         \caption{}
         \label{fig: Corr Functions 1}
     \end{subfigure}
     \hfill
     \begin{subfigure}[b]{0.49\textwidth}
         \centering
         \includegraphics[width=\textwidth]{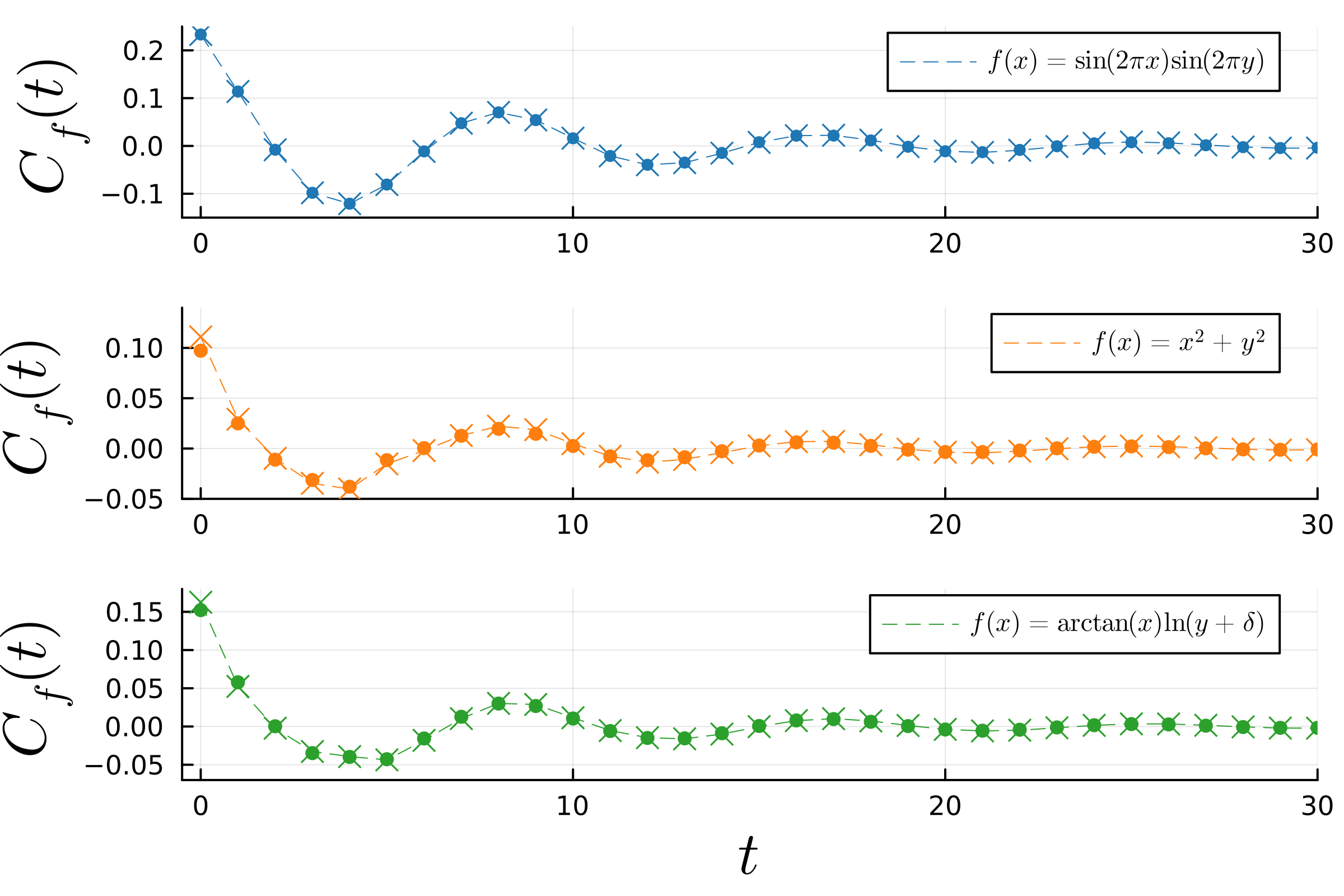}
         \caption{}
         \label{fig: Corr Functions 2}
     \end{subfigure}
        \caption{Panel (a): Correlation functions associated with the eigenfunctions $\varphi_k$. Solid lines refer to direct numerical simulations, dots to the corresponding pure exponential decay. Correlation functions have been normalised by their initial value for scale. Panel (b): Correlation functions (not normalised) for different observables of the system. The regularisation parameter $\delta$ is as in the previous section, $\delta = 0.1$. Crosses correspond to direct numerical simulations, dots to the spectral decomposition. Dashed lines are introduced for visualisation purposes.}
        \label{fig: Correlation functions 2D Map}
\end{figure}

We consider the stochastic chaotic two-dimensional map defined on $\mathcal{M} = \mathbb{T}_{[0,1)}^2 $ generated by the following equations
\begin{align}
\label{eq: Generalised Arnold}
\begin{pmatrix}
    x_{n+1} \\
    y_{n+1} 
\end{pmatrix} = \boldsymbol{\mathcal{A}} \begin{pmatrix}
    x_n \\ y_n
\end{pmatrix} + \frac{1}{\pi} \zeta(x_n+y_n;\mu) \begin{pmatrix}
    1 \\ 1
\end{pmatrix} + \sigma \boldsymbol{\eta} \mod 1,
\end{align}
where 
\begin{equation}
\label{eq: Arnold matrix}
    \boldsymbol{\mathcal{A}} = \begin{pmatrix}
        2 & 1 \\ 1 & 1
    \end{pmatrix},
\end{equation}
and $\zeta : [0,1) \to \mathbb{R} $ is a nonlinear function given by
\begin{equation}
    \zeta(s;\mu) = \arctan \left( \frac{|\mu| \sin\left(2\pi s - \alpha \right)}{1- |\mu| \cos \left(2\pi s - \alpha \right)} \right),
\end{equation}
where $\mu = |\mu|e^{i \alpha} \in \mathbb{C}$ is a parameter with $|\mu| < 1$ and phase defined in the interval $\alpha \in [-\pi,\pi)$. We consider here, $|\mu| = 0.88$ and $\alpha = - 2.4$. The stochastic source of the dynamics is prescribed by $\boldsymbol{\eta}$, a two-dimensional vector of independent Gaussian variables with vanishing mean and unitary variance. In its deterministic version, $\sigma = 0$, this map has been introduced in ~\cite{Slipantschuk2020} as a generalisation of the well-known Arnold's cat map which is obtained when setting $\mu =0$ in the above equations. The Arnold's cat map belongs to the class of Anosov maps, a subset of Axiom A dynamical systems corresponding to highly chaotic, mixing and structurally stable dynamical systems. The cat map represents the paradigmatical example of a chaotic area-preserving system. In other terms, its associated invariant measure is the Lebesgue measure, $\mathrm{d}\mu_0(\mathbf{x}) =\mathrm{d}\mathbf{x}$. The introduction of the nonlinear function $\zeta$ preserves the Anosov property but makes the physical invariant measure of the system singular with respect to Lebesgue, see Panel (a) of Figure \ref{fig: Invariant Measure 2d Map} for a representation of the underlying chaotic fractal attractor.
We here consider a low noise intensity $\sigma = 0.01$, corresponding to an invariant measure being concentrated around the deterministic attractor, see Panel (b) of Figure \ref{fig: Invariant Measure 2d Map}. The authors in ~\cite{Slipantschuk2020} have shown that, for the deterministic version of the map, EDMD yields eigenvalues that correspond to the correct deterministic Ruelle-Pollicott resonances when the natural choice of a Fourier dictionary is made. We here push forward such results and verify firstly that EDMD can be used to spectrally reconstruct the mixing properties of the system. Secondly, we provide evidence that also the response operators can be estimated, through the FDT, using the spectral properties obtained with EDMD. The invariant density  $\rho_0(x)$ is far from being uniformly smooth on the whole phase space $\mathcal{M}$, hindering, at a practical level, the application of the FDT and the EDMD ability to capture the properties of $\rho_0$. 
We fix a Fourier dictionary $\boldsymbol{\Psi}(\mathbf{x}) =\{\psi_\mathbf{k} \}\coloneqq\{  e^{i 2 \pi \mathbf{k}\cdot \mathbf{x}} \}_{\mathbf{k}}$ where $\mathbf{k} = (k_x,k_y)$ with $k_x =  - K_{max}, \dots , 0 , \dots K_{max}$ and similarly for  $k_y$. As done in the previous section, we first assess that the eigenvalues $\nu_i$ of the Kolmogorov matrix $\mathbf{K}$ and the eigenfunctions $\varphi_i(\mathbf{x})$ obtained through EDMD are consistent with the spectral decomposition  \eqref{eq: spectral decomposition correlation function} by checking that $C_{\varphi_i}(t) = c |\nu|^t_i$ with $c$ being a constant, see Panel (a) of Figure \ref{fig: Correlation functions 2D Map}.
Panel (b) instead shows the spectral reconstruction of the mixing properties of generic observables $f$, not necessarily in the span $\mathcal{F}_N$ of the dictionary. We observe quite good agreement with the direct numerical simulations.
\\\\
This map exhibits interesting response properties. Contrary to the previous example, a uniform perturbation $\mathbf{X}(\mathbf{x}) = 1$ would result in a vanishing linear response, which we have verified  (not shown here) both with direct response experiments and with the spectral decomposition. This allows us to verify whether the spectral reconstruction methodology can capture the response to space-dependent perturbations. 
We provide the results for two classes of perturbations. We first consider a sinusoidal perturbation $\mathbf{X}_1(\mathbf{x}) = \sin(2 \pi (x-2y)) (\hat{x} + \hat{y}) $, where $\hat{x}$ ($\hat{y}$) represents the unitary versor in the direction $x$ ($y$). This perturbation profile lies in $\mathcal{F}_N$ and can be fully captured by the dictionary. We thus consider a more complicated type of profile corresponding to perturbing the element $ \boldsymbol{\mathcal{A}}_{11} \to \boldsymbol{\mathcal{A}}_{11} + \varepsilon T(t) $ of the matrix \eqref{eq: Arnold matrix}. This perturbation is associated to a field $\mathbf{X}_2(\mathbf{x}) = x \hat{x} \notin \mathcal{F}_N$.   
\\
As a result of $\rho_0$ being concentrated in narrow areas of the phase space corresponding to the underlying deterministic attractor, the estimation of response properties using \eqref{eq: spectral Green 1 version} as done in the previous section is not reliable. We here then use the alternative procedure given by equation \eqref{eq: spectral Green 2 version}.  We first decompose the observables $\eta_i(\mathbf{x})= \mathbf{X}^i(\mathbf{x}) \rho_0$ into dictionary as $\eta_i = \sum_{\mathbf{k}}\eta^{\mathbf{k}}_i \psi_\mathbf{k}$ and then evaluate the action of the perturbation operator similarly to the previous section as 
\begin{equation}
    \gamma(x) = \mathcal{L}_p \rho_0 = \sum_{i} \partial_i \eta_i =\sum_{\mathbf{k}} \sum_{i} 2 \pi i k_i \eta_i^{\mathbf{k}} \psi_{\mathbf{k}} \coloneqq \sum_{\mathbf{k}} h_{\mathbf{k}} \psi_{\mathbf{k}}.
\end{equation}
The coefficients $\gamma_i$ of the decomposition of the observable $\gamma$ onto the Kolmogorov eigenfunctions $\varphi_i(\mathbf{x})$ can then be obtained using \eqref{eq: from dic to Koopman}. Since the dictionary is given by an orthonormal basis, $\mathbf{G}^{Leb}_{ij} = \delta_{ij}$ and the Green's function is readily estimated using \eqref{eq: spectral Green 2 version} and \eqref{eq: scalar product with respect to Lebesgue}. 
\begin{figure}
     \centering
     \begin{subfigure}[b]{0.49\textwidth}
         \centering
         \includegraphics[width=\textwidth]{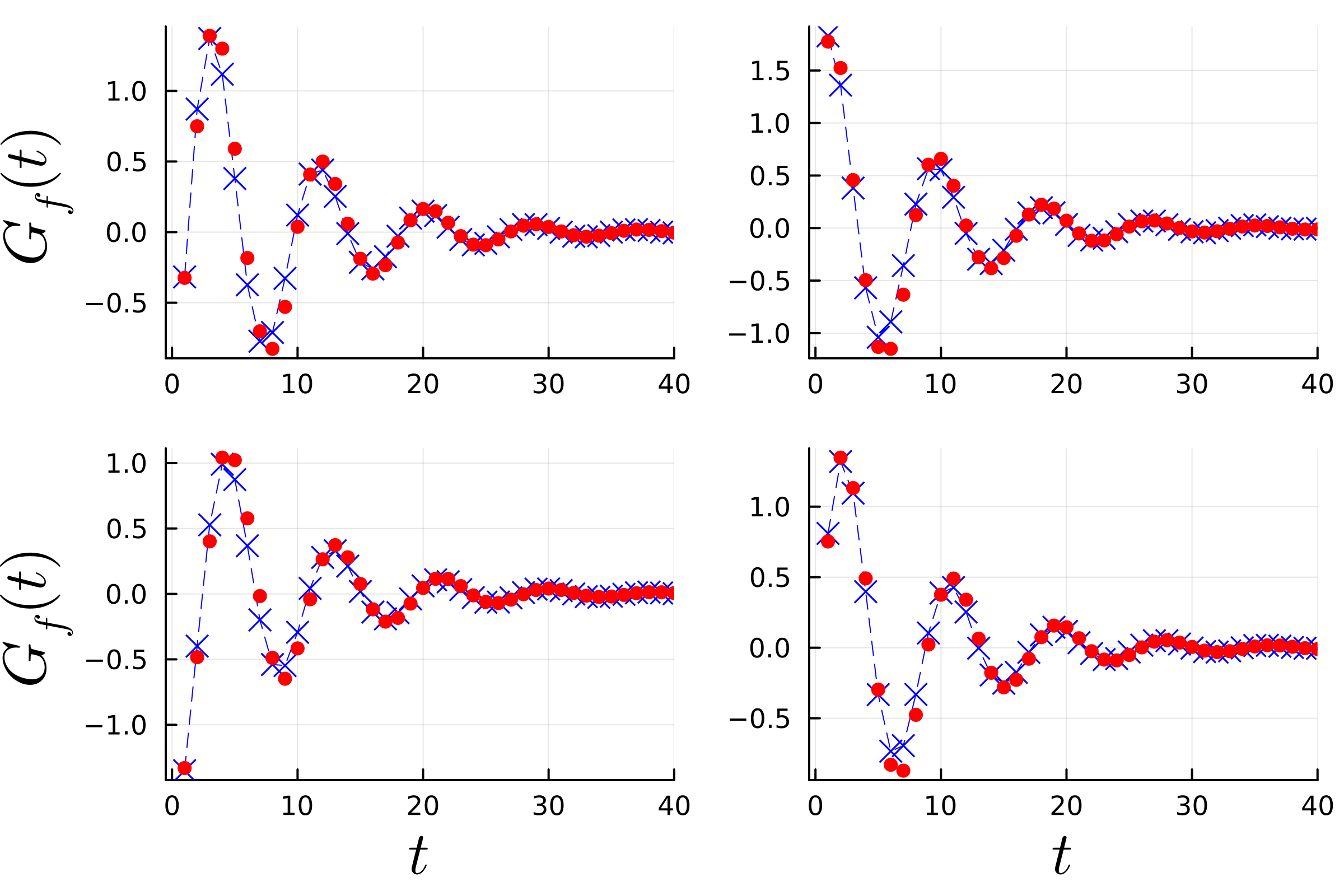}
         \caption{}
         \label{fig: Fourth 2}
     \end{subfigure}
\hfill
 \begin{subfigure}[b]{0.49\textwidth}
         \centering
    \includegraphics[width=\textwidth]{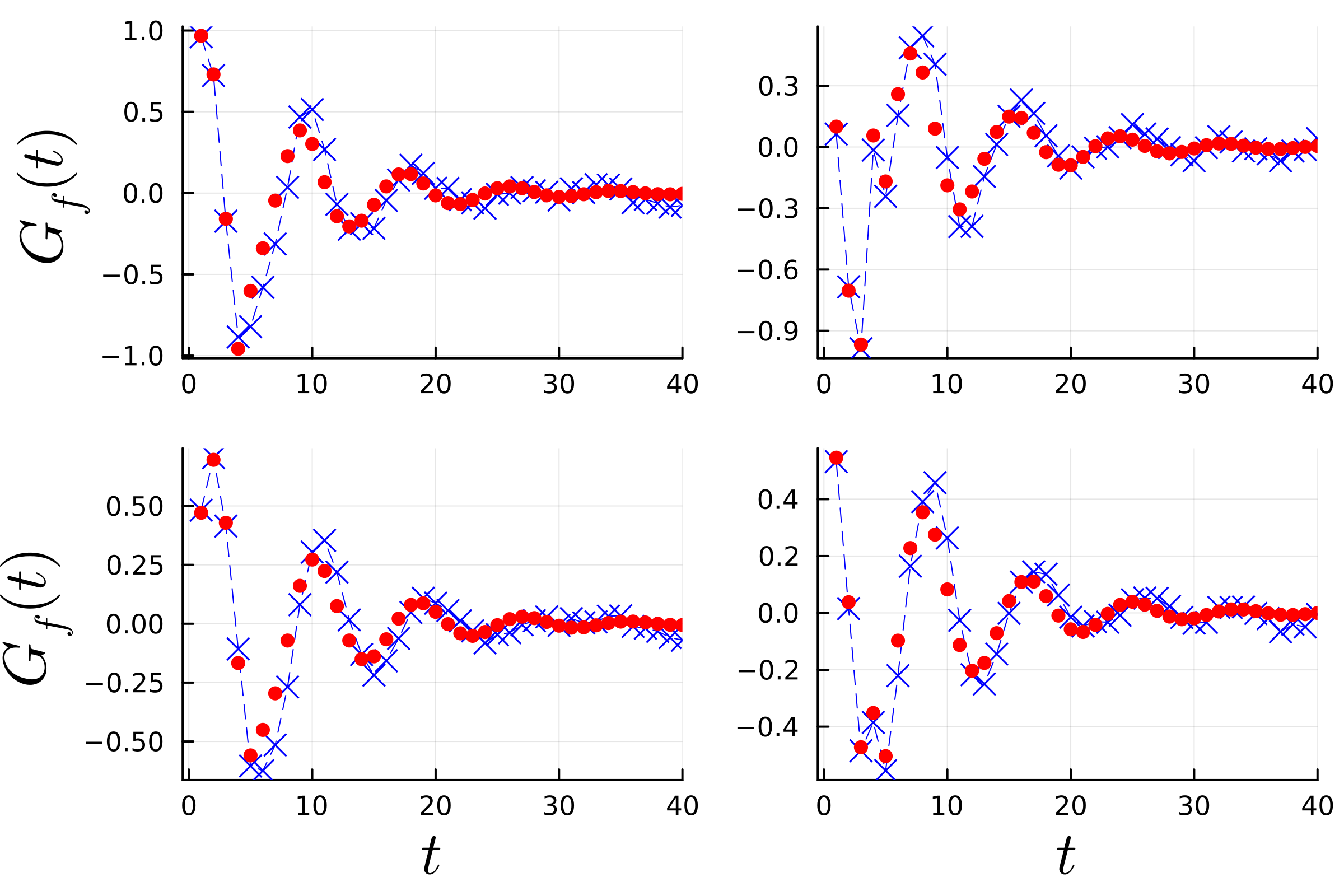}
         \caption{}
         \label{fig: Fourth 3}
     \end{subfigure}
        \caption{Green's functions associated with different observables (from top left to bottom right $f(x,y) = \sin(2\pi(x+y))$, $f(x,y) = \cos(2\pi(x+y))$, $f(x,y) = \sin(2\pi x)\cos(2\pi y)$, $f(x,y) = \cos(2\pi x)\cos(2\pi y)$).  Panel (a): Response to the perturbation field $\mathbf{X}_1(\mathbf{x}) $. Panel (b): Response to a perturbation $\boldsymbol{\mathcal{A}}_{11} \to \boldsymbol{\mathcal{A}}_{11} + \varepsilon T(t)
         $. Crosses correspond to direct numerical response experiments, dots to the spectral decomposition. Dashed lines are introduced for visualisation purposes. }
        \label{fig: Fourth}
\end{figure}
Figure \ref{fig: Fourth} shows a very good agreement between the spectrally reconstructed Green's function and direct numerical response experiments. Here we have used $K_{max} = 8$ and a trajectory of length $M=2\times 10^6$ for the evaluation of the Kolmogorov properties using EDMD. It is remarkable that, even for a relatively low number of dictionary functions, the EDMD spectral reconstruction can pick up the key features of the response of the quite peaked reference state $\rho_0$. To fully resolve the properties of the invariant measure, we have here estimated $\rho_0$ from a long trajectory $M= 10^7$ but have verified that the reconstruction of the $G_f(t)$ is stable for a quite wide range $M \in [10^5 , 10^8]$. We have also considered here Green's functions associated with periodic observables that belong to $\mathcal{F}_N$ so that the discrepancies in the estimation of the $G_f(t)$ are to be attributed to either quadrature problems in the Galerkin approximation given by EDMD or by a not perfect representation of $\eta_{i}(\mathbf{x}) = \mathbf{X}^{i}(\mathbf{x}) \rho_0 \notin \mathcal{F}_N$ by the dictionary. The comparison between panel (a) and (b) of Figure \ref{fig: Fourth} shows that the response associated with $\mathbf{X}_1 \in \mathcal{F}_N$ is better captured than the response to $\mathbf{X}_2 \notin \mathcal{F}_N$. Similarly to the analysis shown in Figure \ref{fig:Robustness}, a larger dictionary (associated with a longer trajectory $M$) results in a better estimate of the Green's function. We have here shown the results for $K_{max} = 8$ to highlight biases and limitations of the spectral reconstruction. We also remark that, in most applications, the physically relevant quantity is not the Green's function per se but its convolution $( G_f \star T)(t)$ with specific time dependent forcings.
 \begin{figure}
     \centering
     \begin{subfigure}[b]{0.49\textwidth}
    \includegraphics[width=\textwidth]{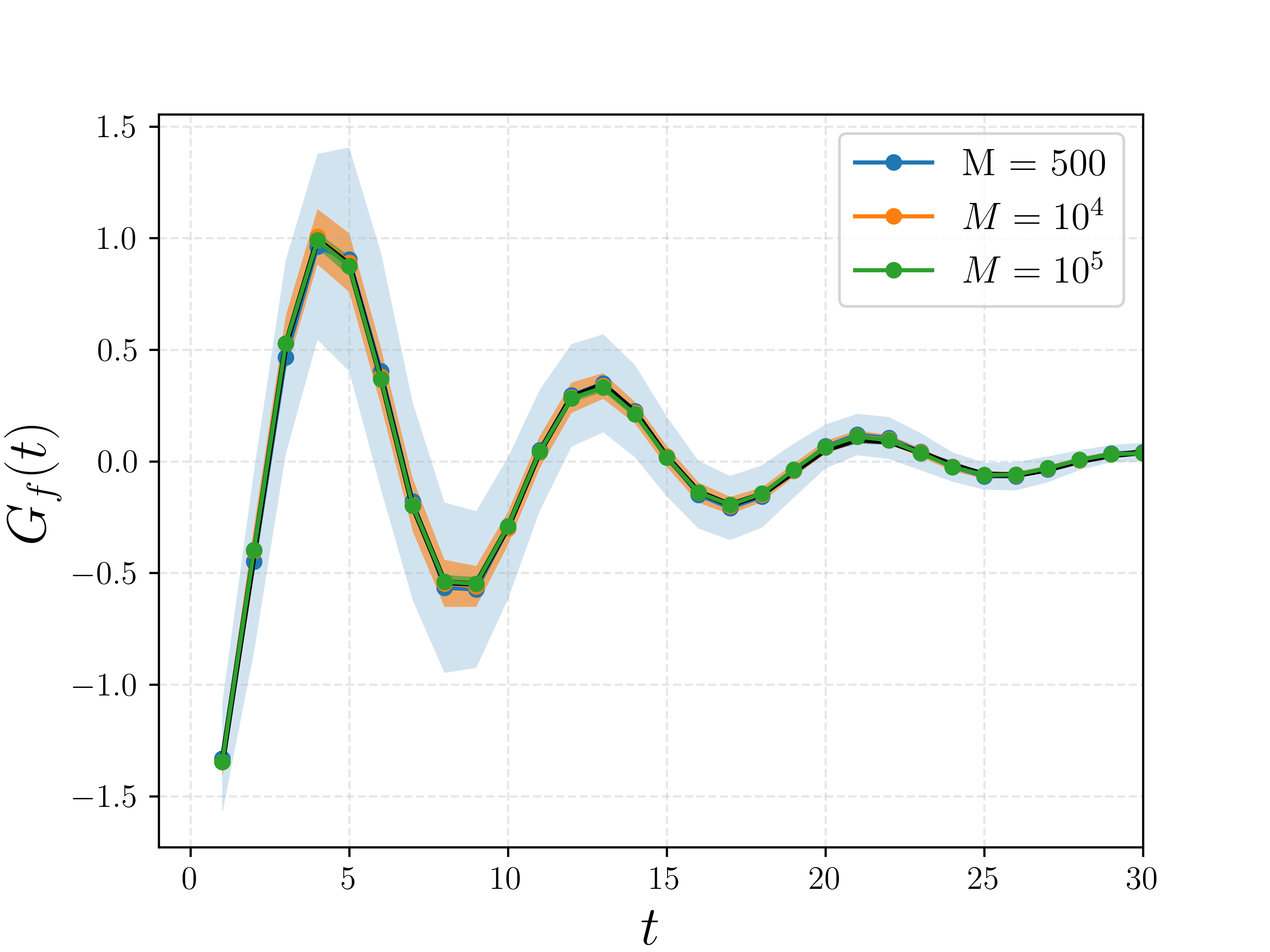}
    \caption{}
     \end{subfigure}
     \hfill
     \begin{subfigure}[b]{0.49\textwidth}
         \includegraphics[width=\textwidth]{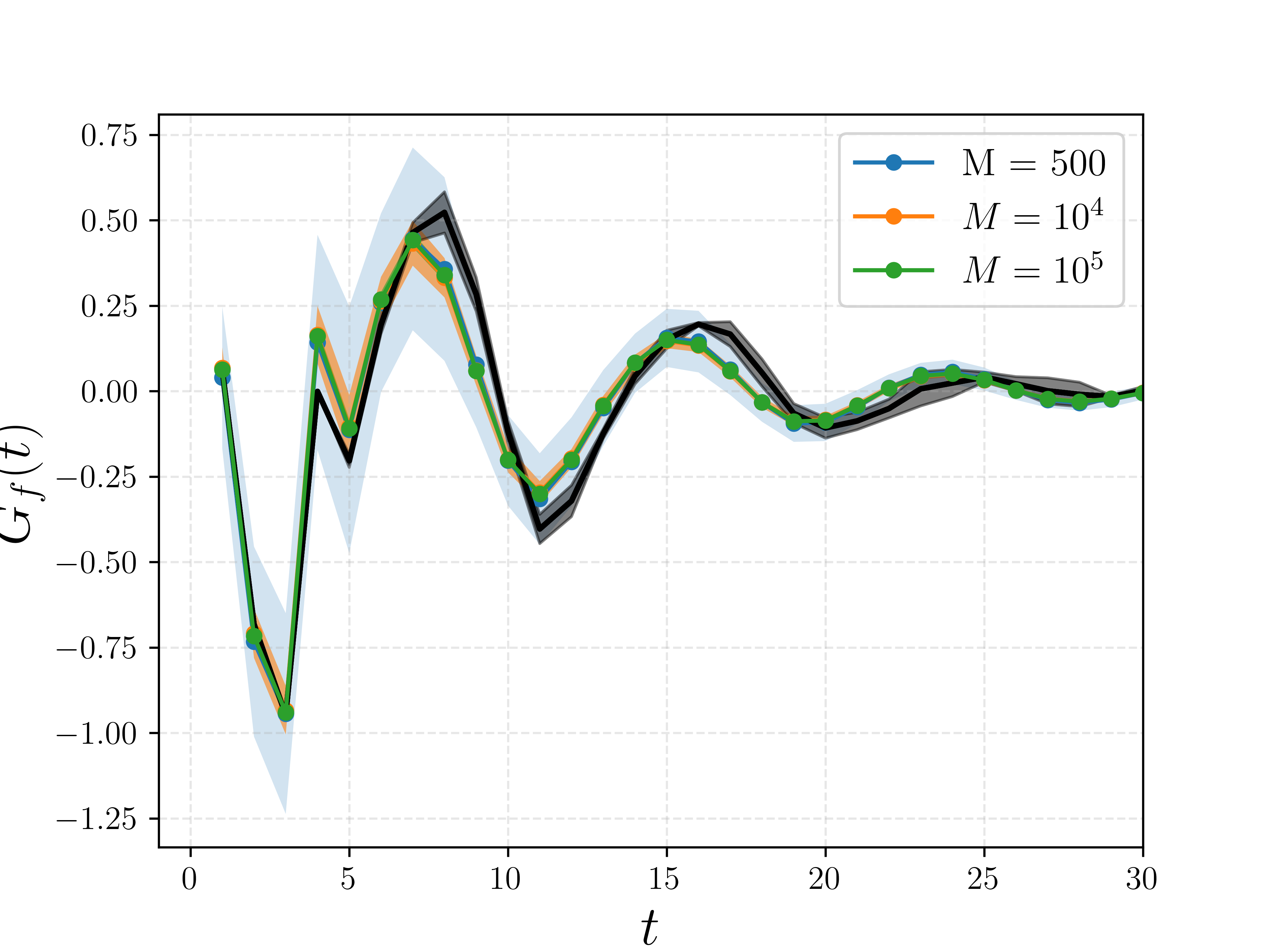}
         \caption{}
     \end{subfigure}
     \caption{Green's functions obtained by estimating projections of $\Gamma$ from trajectory data. Panel (a): Response to the perturbation field $\mathbf{X}_1(\mathbf{x})$ of observable $f(x)=\sin(2\pi x)\cos(2\pi y)$. Panel (b): Response to a perturbation $\boldsymbol{\mathcal{A}}_{11} \to \boldsymbol{\mathcal{A}} + \varepsilon T(t)$ of observable $f(x)= \cos(2\pi(x+y))$. The shaded area corresponds to one standard deviation range at each point in time, estimated from an ensemble of $100$ independent realisations. The black line is the Green's function evaluated from direct response experiments on an ensemble of $10^6$ realisations. The black shaded area corresponds to uncertainty in the Green's function evaluated as the standard deviation across response experiments with different perturbation amplitude $\varepsilon$. Here, spectral properties of $\mathcal{K}_0$ and projections of $\Gamma$ are evaluated on the same data.}
     \label{fig: New method}
 \end{figure}
 A more data-efficient spectral reconstruction of the response can be obtained by using Method 3, described at the end of section \ref{sec: spectral reconstruction}. To test the performance and robustness of this method for varying trajectory lengths, we perform an ensemble of $100$ independent experiments and track the average and standard deviation of the estimate of the Green's function across the ensemble. Here, spectral properties of the Kolmogorov operator $\mathcal{K}_0$ and projections of the response observable $\Gamma$ are obtained from the same trajectory data, as opposed to the previous Methods. Figure \ref{fig: New method} shows that on average this reconstruction method is able to accurately estimate the response with as little as $M=500$ data points. As expected, a larger data set corresponds to a smaller standard deviation of the estimate. It is remarkable that this method achieves very good accuracy with such a small data set, when, in order to estimate the Green's function from direct numerical experiments (black line in the Figure), one needs to consider an ensemble of $10^6$ response experiments. To regularise the EDMD procedure we have used a truncated singular value decomposition $\mathbf{G}=\mathbf{U}\mathbf{S}\mathbf{U}^\dagger$ and retained only the first $r$ singular values $\sigma_i = S_{ii}$ $i=1,\dots,N$ for which $\sigma_i / \sigma_1 > 10^{-4}$.
{\color{black}{
\subsection{Double-well potential}
\label{sec: double well}
Finally, we consider a stochastic two-dimensional gradient flow with a double-well potential. While this system is not chaotic, the double-well provides a useful test case for EDMD, as analytic expressions for the invariant measure and related quantities may be obtained directly, and the Kolmogorov eigenfunctions may be visualised and have a level of interpretability. Furthermore, in contrast to the discrete maps studied so far, this is a continuous-time example.
 We study the two-dimensional SDE 
\begin{equation} \label{eq: gradient sde}
    \mathrm{d}\mathbf{x} = -\nabla V(\mathbf{x})\mathrm{d}t + \sigma \mathrm{d}\mathbf{W}_t, \quad V(x,y) = (x^2-1)^2+y^2,
\end{equation}
where $\mathbf{W}_t = ({W}_{t,x},{W}_{t,y})$ is a vector whose components are two independent Wiener processes and $\nabla$ indicates the customary two--dimensional gradient operator. A Koopman analysis of the stochastic double well system has been pursued before \cite{WilliamsKevrekedis2015, klus2016NumericalApproximation}, and indeed, for this example, we consider an identical system to that in \cite{klus2016NumericalApproximation}, so that we consider an additive noise perturbation and select $\sigma = 0.7$. In contrast with that example, however, we perform EDMD by taking a single long trajectory so that our snapshot data is distributed with respect to the invariant measure as in the previous two sections. Beginning with an initial condition taken randomly from the region $x,y \in (-1.5,1.5)$, we employ an Euler-Maruyama method \cite{kloeden-platen-92} to integrate from $t=0$ to $t=5 \times 10^5$ with a step size $\delta t = 10^{-3}$.
\\
\begin{figure}
    \centering
    \includegraphics[angle = 270,width=0.8\textwidth]{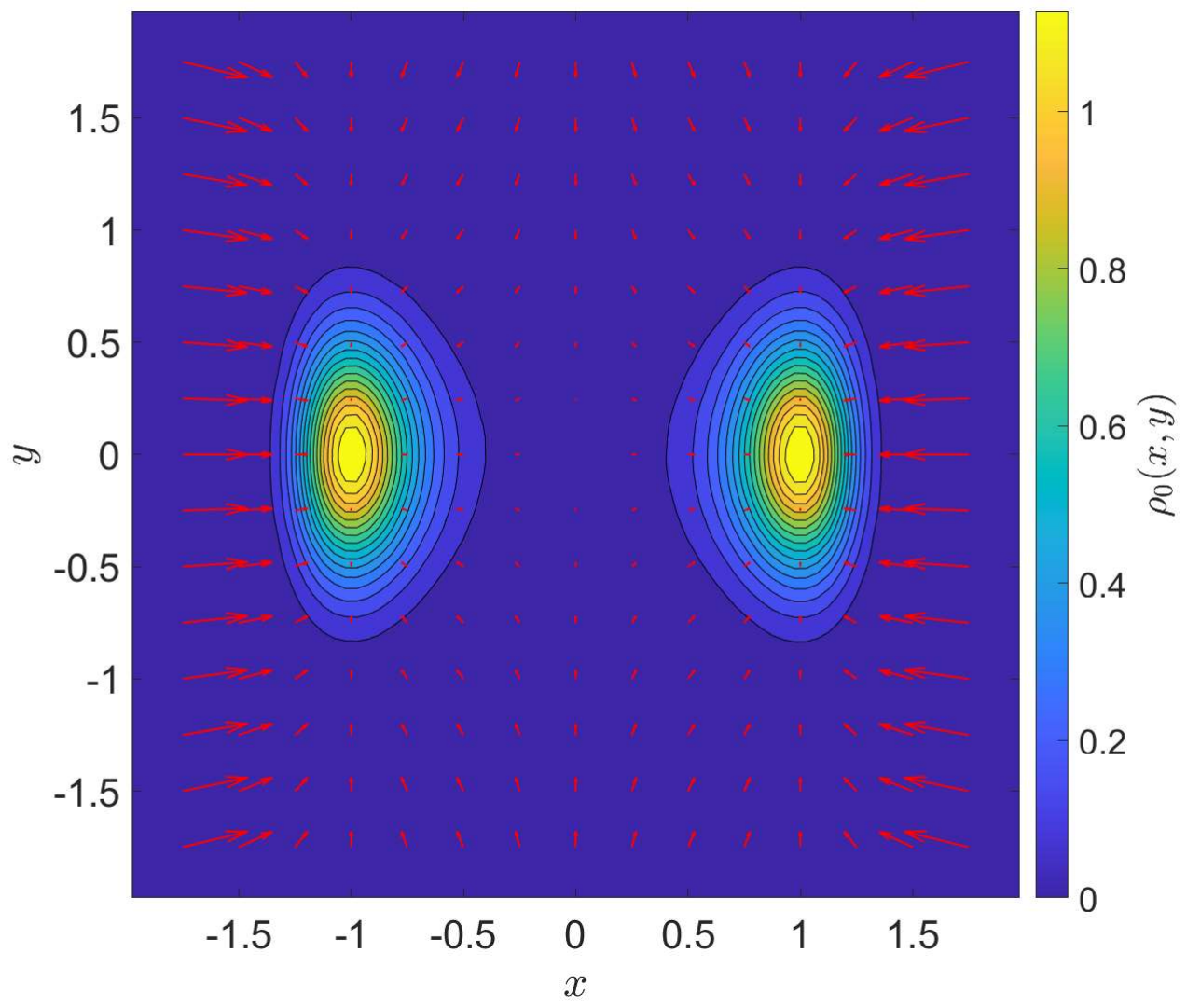}
    \caption{Invariant density of the stochastic double well system, with deterministic dynamics $\mathbf{F}(\mathbf{x})$ overlaid as arrows. }
    \label{fig: Invariant Density 2Well}
\end{figure}
As a consequence of the gradient structure of \eqref{eq: gradient sde}, the system preserves the microscopic detailed balance condition and is thus a classical example of an equilibrium statistical mechanical system. Many important theoretical results can be obtained for this system. Firstly, it is possible to prove that the system converges exponentially fast to the invariant Gibbs measure
\begin{equation} \label{eq: inv dens db well}
    \rho_0(\mathbf{x}) = \frac{1}{Z}e^{-2V(\mathbf{x})/\sigma^2},
\end{equation}
with normalization constant $Z = \int e^{-2V(\mathbf{x})/\sigma^2} \mathrm{d}\mathbf{x}$. Secondly, as a consequence of the microscopic reversibility of the gradient dynamics, it is possible to prove that the Kolmogorov operator is self adjoint in $L^2_{\mu_0}$ \cite{pavliotisbook2014}. As a result, the stochastic resonances $\lambda_j$'s describing the decay towards the statistical equilibrium are all real and no oscillations can be observed in the system, as opposed to the previous two nonequilibrium examples.
\\
 As may be seen from Figure \ref{fig: Invariant Density 2Well}, the trajectory spends most of the time localised in either well, so that there is a time-scale separation between inter and intra-well dynamics. 
\\
\begin{figure}
    \centering
    \includegraphics[angle = 270,width=\linewidth]{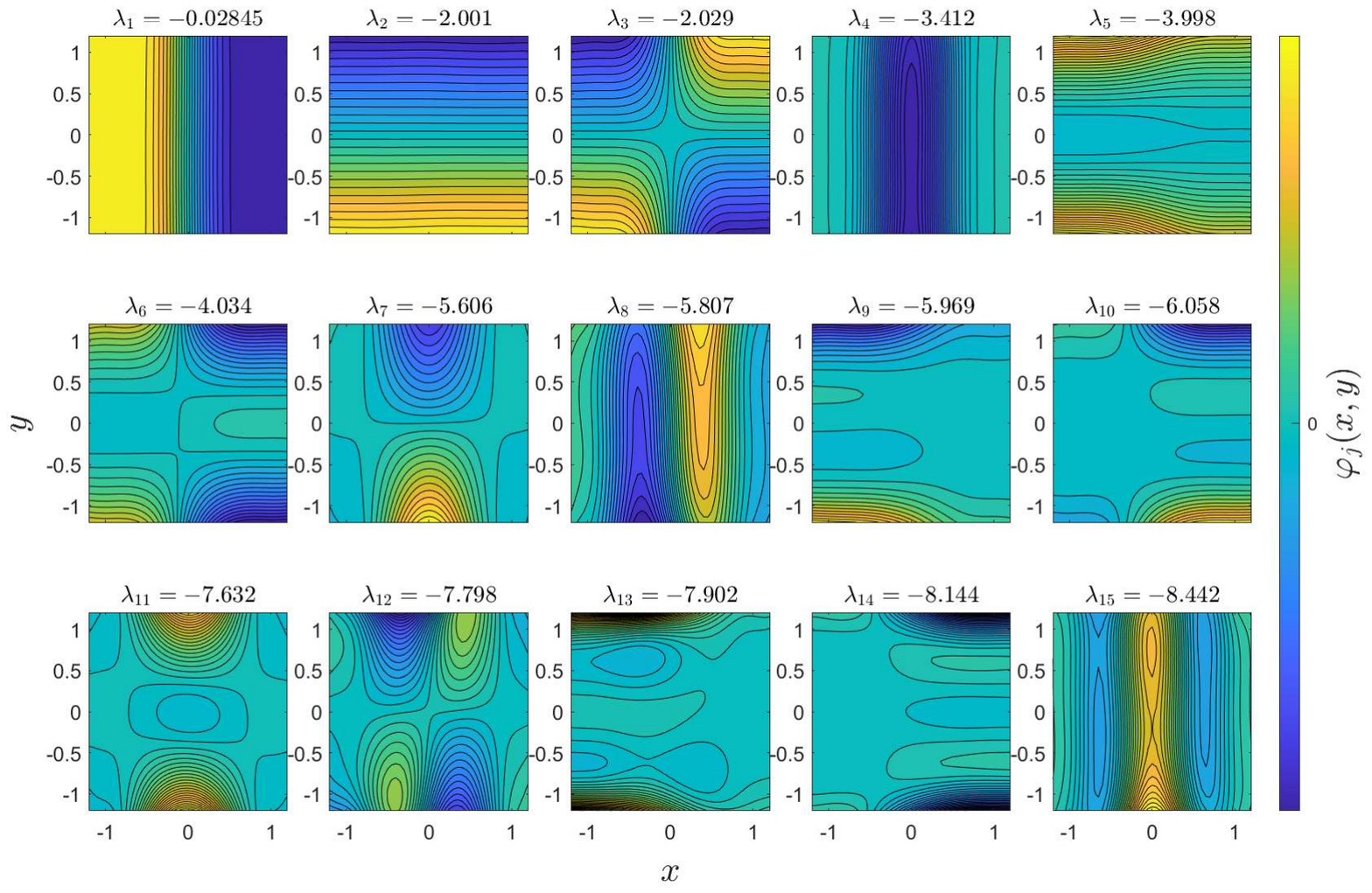}
    \caption{The leading 15 subdominant eigenfunctions of the Kolmogorov operator for the stochastic double well, labelled by their corresponding eigenvalue. The first subdominant eigenvector succinctly describes the left-right transition between the two potential wells. As eigenfunctions are defined up to an arbitrary scaling factor, only the zero of the colour scale is labelled.}
    \label{fig: eigenfns dbwell}
\end{figure}
As the deterministic dynamics $\mathbf{F}(\mathbf{x}) = -\nabla V(\mathbf{x})$ is expressed in terms of polynomials in $x$ and $y$, the natural choice of dictionary for EDMD is one of polynomials. In keeping with \cite{klus2016NumericalApproximation}, we choose monomials up to order 10 as basis functions, yielding a dictionary with $66$ terms:
\begin{equation}
    \boldsymbol{\Psi}(\mathbf{x}) = \left(1,x,y,x^2,xy,y^2,\dots,xy^9,y^{10}\right).
\end{equation}
For the snapshot data, we filter the trajectory and discard an initial transient so that we consider $M = 10^7$ pairs of points separated by time $\Delta t = 0.05$, corresponding to a trajectory of length $T= 5\times 10^5$ time units. Here, we have considered a very long trajectory to obtain robust and clear pictures of the Kolmogorov eigenfunctions. We anticipate that the spectral reconstruction of the response is stable over a large range of trajectory lengths. As before, we calculate the eigenvalues and eigenvectors of the Kolmogorov matrix $\mathbf{K}$ given in Eq. \ref{eq: Kolmogorov matrix}. 

Since this is a \textit{bona fide} equilibrium system, as mentioned above, we adopt here the Hermitian DMD algorithm \cite{Drmac2022,Drmac2024}, which basically amounts to imposing hermitianity of the $\mathbf{A}$ matrix defined in Eq. \ref{matrixA} by setting $\mathbf{A}\rightarrow 1/2 (\mathbf{A}+\mathbf{A}^*)$. The good convergence properties of this algorithm have been recently assessed in detail \cite{colbrook2024hermitian}. We also remark that the direct application of the standard EDMD provides almost indistinguishable results, both in terms of eigenvalues and corresponding eigenvectors.

As this is a continuous-time system, the eigenvalues $\nu_j$ depend on our choice of timestep, $\Delta t$, and the eigenvalues $\lambda_j$ of $\mathcal{K}_0$ are the physically relevant quantities. 
The first $15$ of these, sorted in descending order, are printed alongside their corresponding eigenfunctions in Figure \ref{fig: eigenfns dbwell}. The eigenvalues are all negative real numbers, in agreement with the gradient nature of the system. The absolute value of the sub-leading eigenvalue $\lambda_1$ is two orders of magnitude smaller than all the others, and its corresponding eigenfunction is uniform, apart from a sudden but smooth transition from positive to negative values when going from one well to the other. In a physical sense, this is related to the transfer of mass between wells, and we can interpret this as the dominant coarse-grained process in the system \cite{klus2016NumericalApproximation}, which justifies the usually assumed two-state approximations for this class of models.  Intra-well dynamics are encoded in the higher-order eigenfunctions.
\\
\begin{figure}
    \centering
    \begin{subfigure}[b]{0.49\textwidth}
         \centering
    \includegraphics[angle = 270,width=\textwidth]{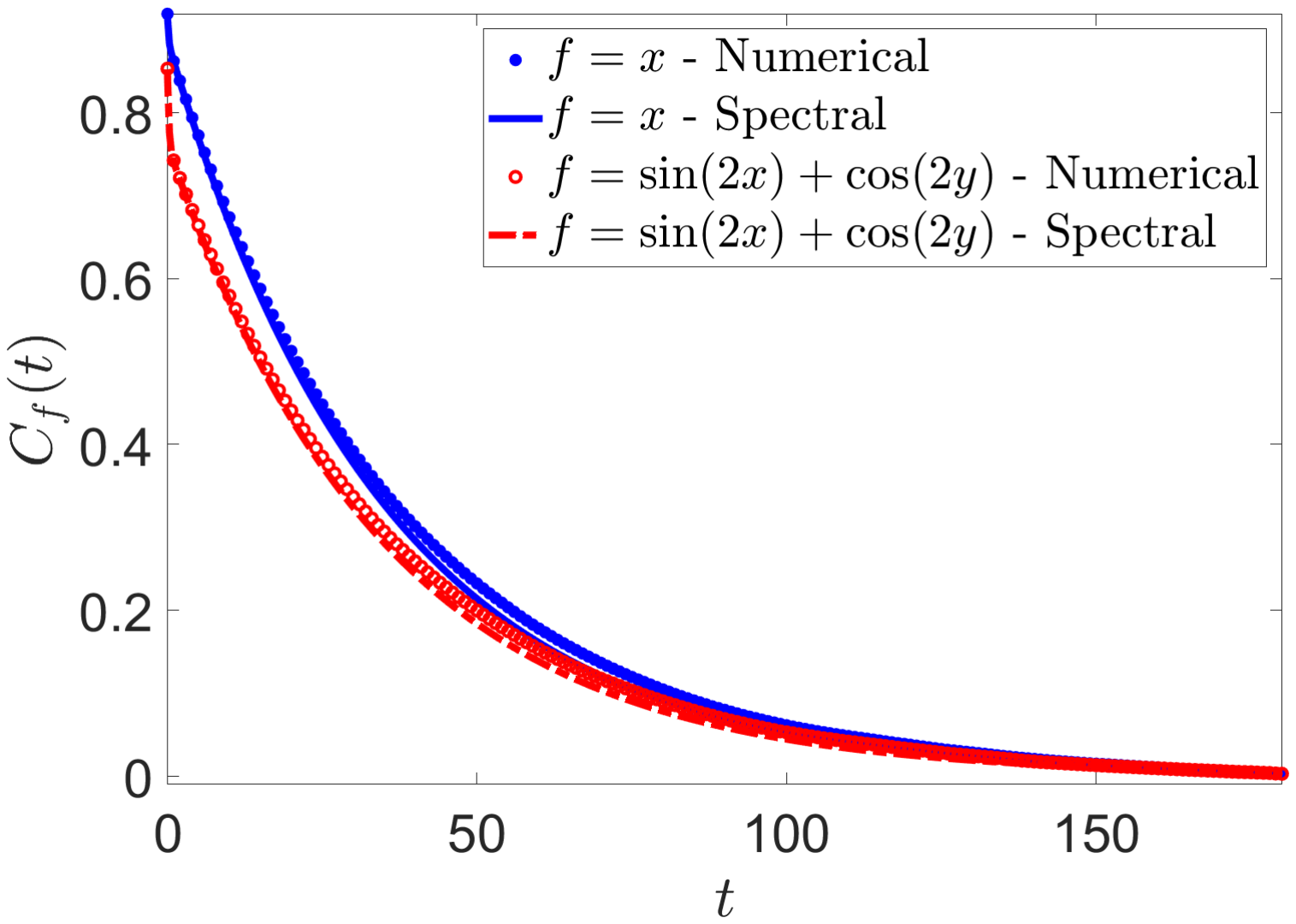}
         \caption{}
         \label{fig: autocorrs 2well odd x}
     \end{subfigure}
     \hfill
     \begin{subfigure}[b]{0.49\textwidth}
         \centering
         \includegraphics[angle = 0,width=\textwidth]{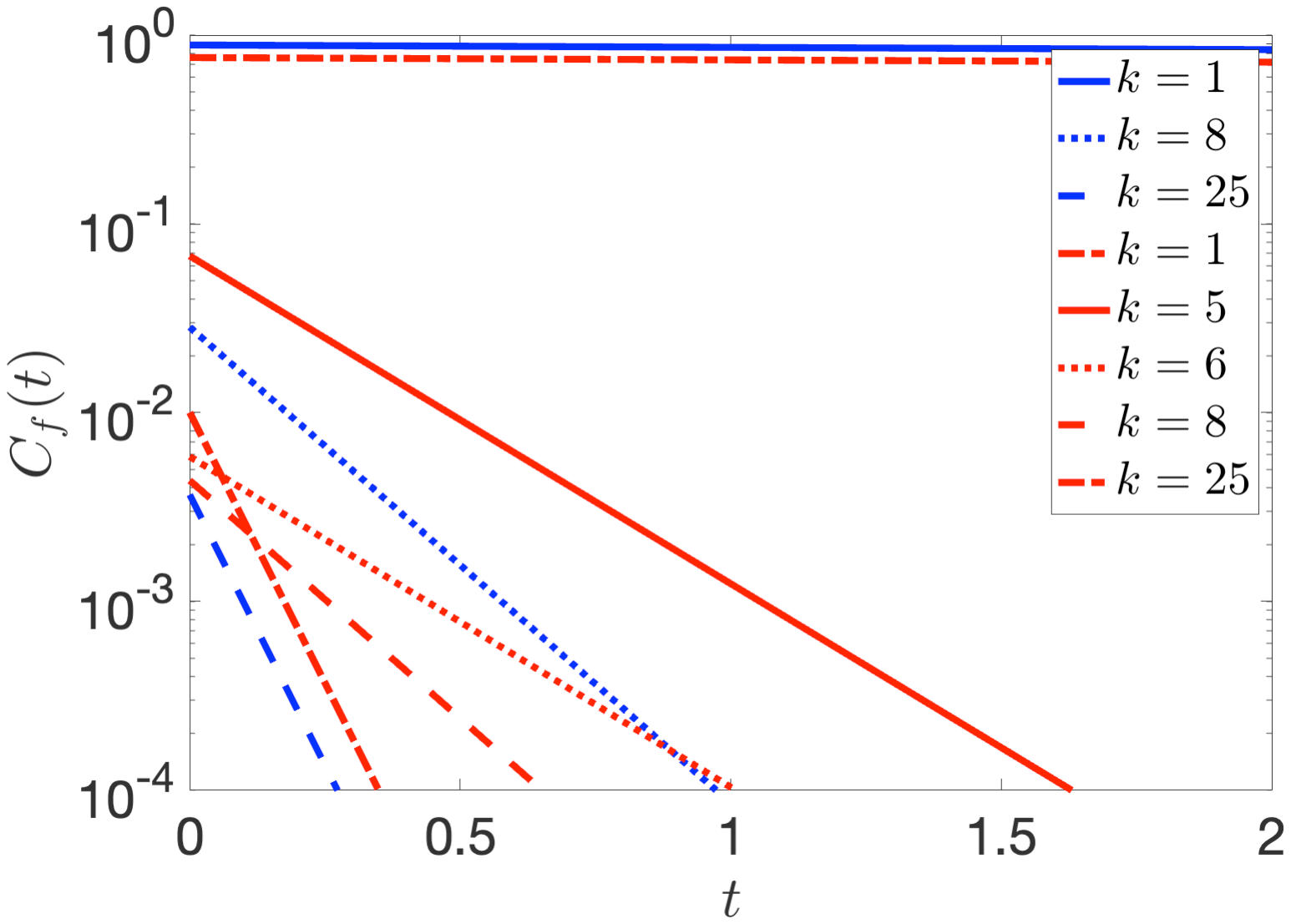}
         \caption{}
         \label{fig: corr decomp 2Well odd x}
     \end{subfigure}
    \caption{Panel (a): Autocorrelation functions of two observables that are odd in $x$. The round markers represent results from direct numerical correlation functions of the observables, while the solid and dashed lines of the same colour correspond to results of spectral decomposition. Panel (b): Decomposition of the correlation functions from (a) into the individual terms in \eqref{eq: spectral decomposition correlation function} with largest magnitude. The terms from the autocorrelation function of $f(x,y) = x$ are plotted as thicker blue lines, while the thinner red lines correspond to terms from the autocorrelation function of $f(x,y) = \sin(2x) +\cos(2x)$. }
    \label{fig: Corr fns 2well odd x}
\end{figure}
\begin{figure}
     \centering
     \begin{subfigure}[b]{0.49\textwidth}
         \centering
\includegraphics[angle = 270,width=\textwidth]{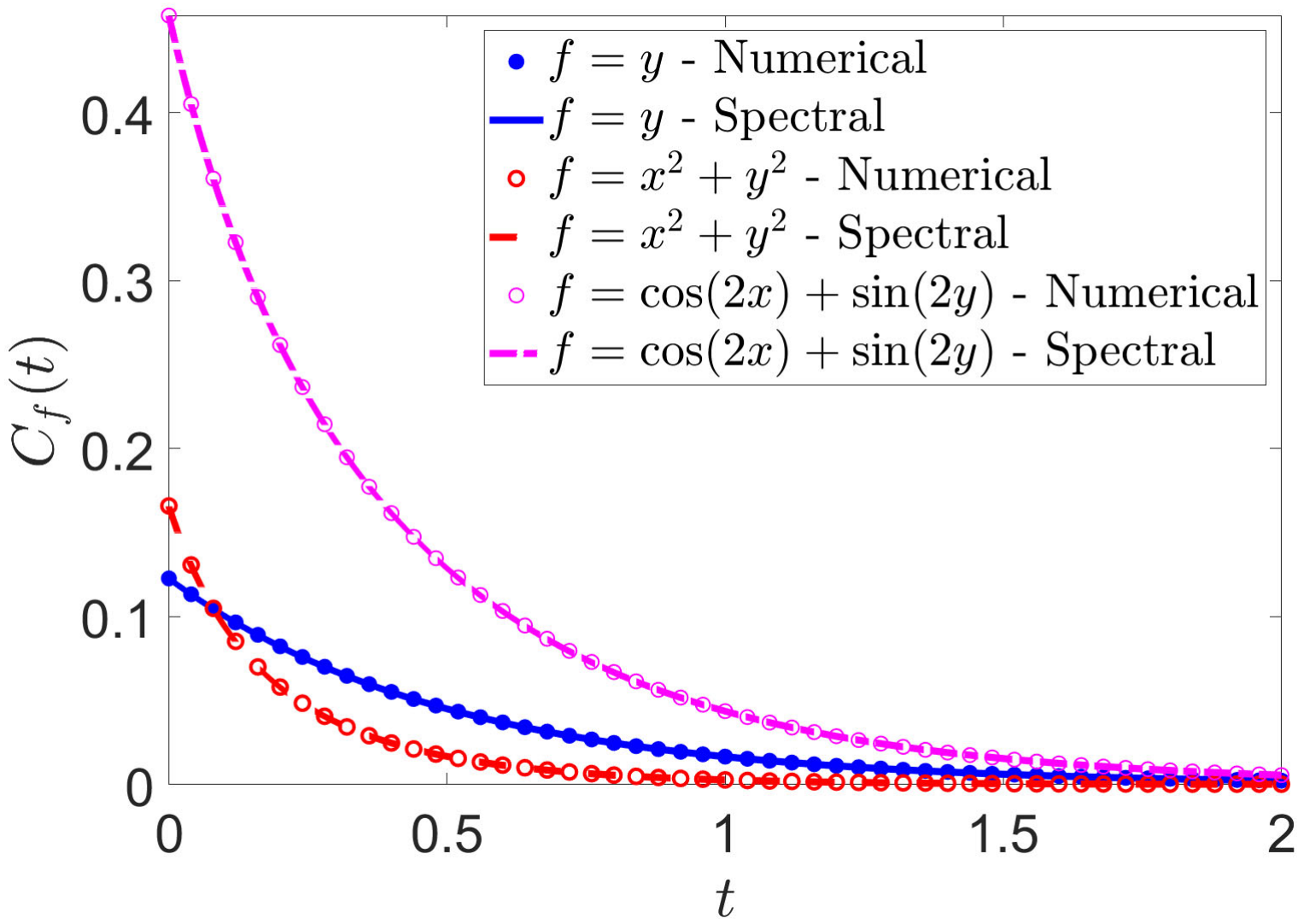}
         \caption{}
         \label{fig: autocorrs 2well even x}
     \end{subfigure}
     \hfill
     \begin{subfigure}[b]{0.49\textwidth}
         \centering
         \includegraphics[angle = 0,width=\textwidth]{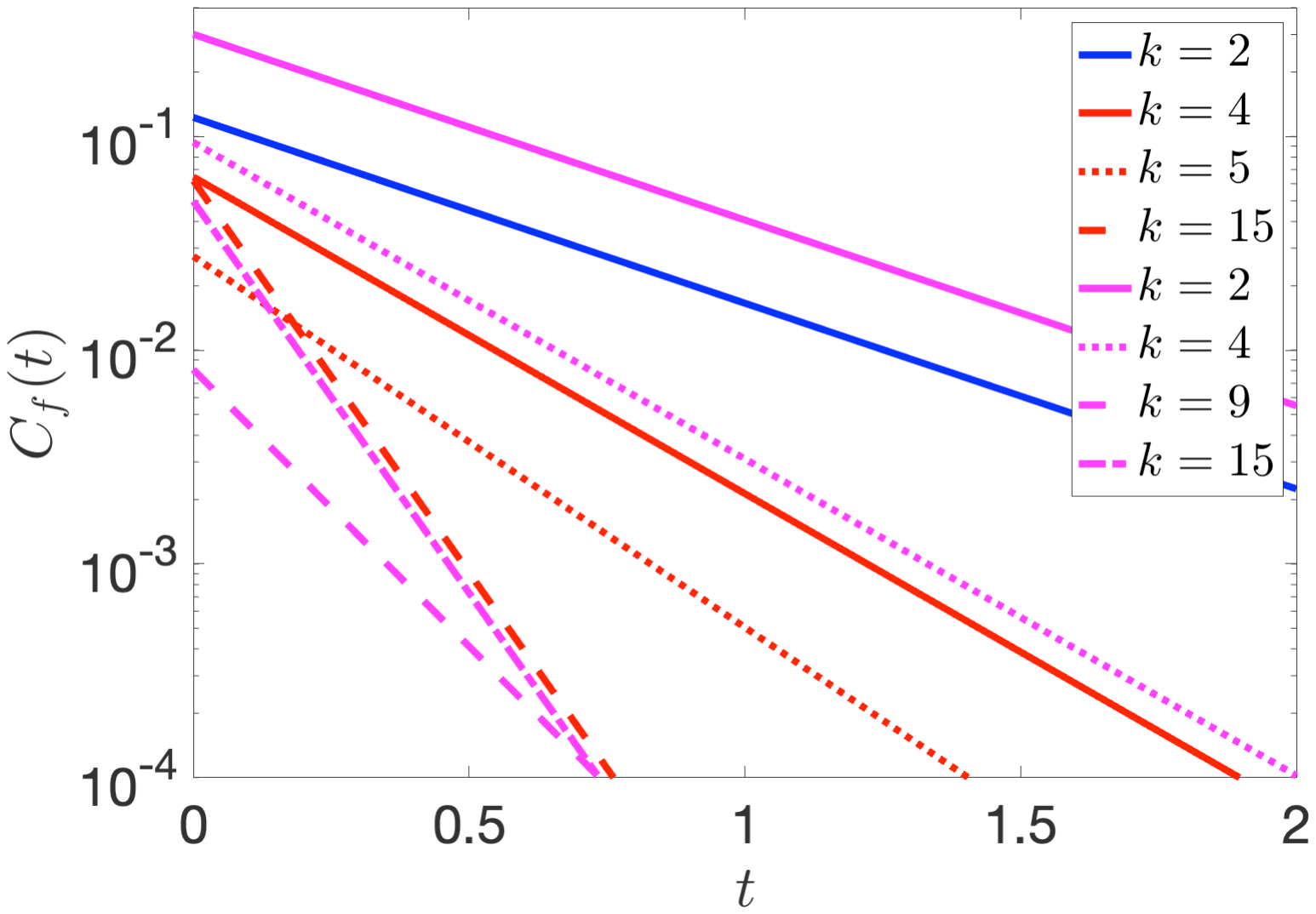}
         \caption{}
         \label{fig: corr decomp 2Well even x}
     \end{subfigure}
        \caption{Panel (a): Autocorrelation functions of three observables that are even in $x$. The round markers represent results from direct numerical correlation functions of the observables, while the solid and dashed lines of the same colour correspond to results of spectral decomposition. Panel (b): Decomposition of the correlation functions plotted in (a) as in Fig. \ref{fig: Corr fns 2well odd x}. 
        The only term from the autocorrelation function of $f(x,y) = y$ shown is that with $k=2$ (blue line), as this term is several orders of magnitude larger than all others at all times. The other lines in this plot all correspond to terms from the autocorrelation function of $f(x,y) = x^2 + y^2$ (red lines) and $f(x,y) = \cos(2x) + \sin(2y)$ (magenta lines).}
        \label{fig: Corr fns 2Well even x}
\end{figure}
As in the previous examples, we decompose observables into a linear combination of these eigenvectors (and the $50$ higher-order modes not shown) to evaluate correlation functions according to Eq. \ref{eq: spectral decomposition correlation function}. Some examples are shown in Figures \ref{fig: autocorrs 2well odd x} and \ref{fig: autocorrs 2well even x}. Once again, the success of the EDMD method is demonstrated by the close agreement of results obtained through spectral decomposition and direct numerical evaluation from trajectory data. As implied by Eq. \eqref{eq: spectral decomposition correlation function}, we see exponential decay of general correlation functions with the absence of oscillations.  

Depending on the choice of the observable, we can expect very different rates of decay of correlation. It is clear that if the observable is not even with respect to $x$, it will have a non-negligible projection over the subdominant mode $\varphi_1$ so that one expects a slow decay of correlations with a dominant time scale given by $1/|\lambda_1|$, see Figure \ref{fig: autocorrs 2well odd x}. 
If the observable is even with respect to $x$, one expects a much faster decay of correlations. Specifically, if the observable is not even with respect to $y$, the dominant time scale will be given by $1/|\lambda_2|\ll 1/|\lambda_1|$. In the case the observable is even with respect to both $x$ and $y$, the dominant time scale of decorrelation is $<1/|\lambda_3|$, see Figure \ref{fig: autocorrs 2well even x}. 

In Figures \ref{fig: corr decomp 2Well odd x} and \ref{fig: corr decomp 2Well even x}, we decompose the autocorrelation functions plotted in the corresponding (a) panels into their constituent terms described in Eq. \ref{eq: spectral decomposition correlation function} 
that have the largest contribution to the sum.

The separation of timescales between inter-well and intra-well dynamics is clearly seen in Figure \ref{fig: autocorrs 2well odd x}. For long times, both correlation functions are dominated by the leading term in the sum, as both functions, having a dominating odd component in $x$, project strongly onto $\varphi_1$. For up to $t\approx 1$, one can see the effect of the faster decaying modes, which are responsible for what on the time scale depicted in Figure \ref{fig: autocorrs 2well odd x} appears as an almost instantaneous decrease of the value of the correlation function. As the rate of the decay of the subdominant mode is two orders of magnitude slower than all higher-order terms, in the log-lin plot depicted in Figure \ref{fig: corr decomp 2Well odd x} it appears as an approximately horizontal line. 

The slow inter-well dynamics are absent in Figure \ref{fig: corr decomp 2Well even x}. As expected, the function $f(x,y) = y$ projects most strongly onto $\varphi_2$, and indeed all other terms in the decomposition given in Eq. \ref{eq: spectral decomposition correlation function} have a negligible role. The leading contribution of the decay of correlation for $f(x,y) = \cos(2x)+\sin(2y)$ is also given by the second subdominant model. Finally, as $f(x,y) = x^2 + y^2$ is even in $y$, it does not project onto the second subdominant mode and has an even faster leading decorrelation rate.
\\
\begin{figure}
     \centering
     \begin{subfigure}[b]{0.49\textwidth}
         \centering
         \includegraphics[angle = 270,width=\textwidth]{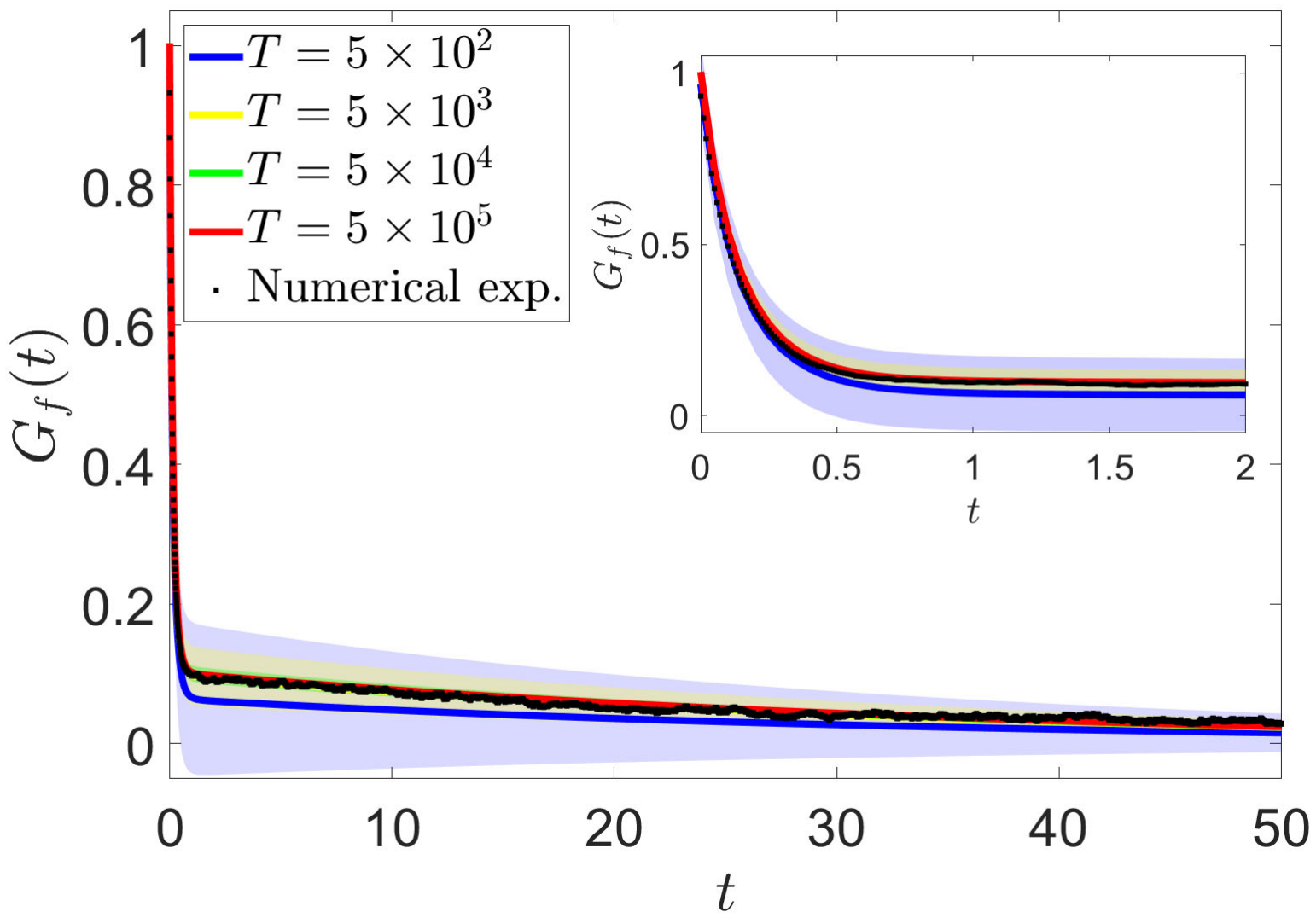}
         \caption{}
         \label{fig: GF 2well unif x }
     \end{subfigure}
     \hfill
     \begin{subfigure}[b]{0.49\textwidth}
         \centering
         \includegraphics[angle = 270,width=\textwidth]{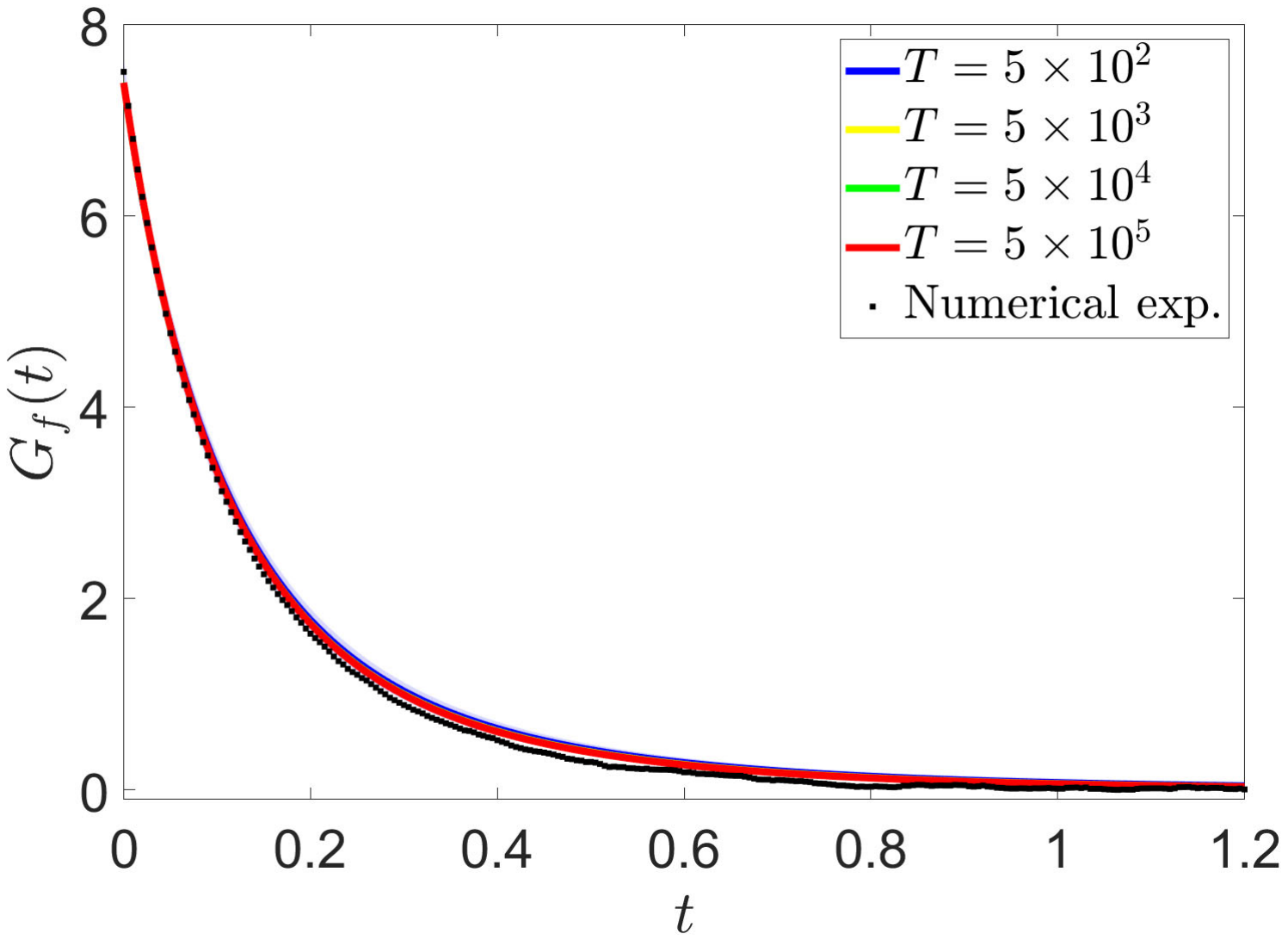}
         \caption{}
         \label{fig: GF 2well radius }
     \end{subfigure}
        \caption{Panel (a): Green's functions measuring the response of the observable $f(x,y) = x$ to the perturbation $\mathbf{X}_1$. Black points correspond to direct numerical response experiments, solid lines to results obtained through spectral decomposition. The inset emphasizes the transition between two regimes of response for $t\approx1$. Panel (b):  Green's functions for $f(x,y) = x^2+y^2$ and $\mathbf{X}_2$. Note the much faster decay rate as compared to Panel (a). Green's functions obtained through the spectral method are averaged over $60$ independent EDMD runs for 4 different trajectory lengths, $T$. The shaded area corresponds to one standard deviation range at each point in time, and is vanishingly small for most values of $T$.}
        \label{fig: GFs 2well}
\end{figure}
To test the linear response properties of the system, we consider two different types of perturbation: a uniform perturbation $\mathbf{X}_1(\mathbf{x}) = (1,0)$ which is equivalent to tilting the potential $V(x,y)$ in the $x$-direction, and a space-dependent perturbation equivalent to varying the distance between the two minima $\mathbf{x}^* = (\pm 1,0)$ of the wells as $\mathbf{x}^* \to \mathbf{x}^*  \pm (\varepsilon T(t),0)$. It is simple to see that such perturbation corresponds to a profile $\mathbf{X}_2(\mathbf{x}) = (4x,0)$.  The Green's functions associated with these perturbations are then correlation functions involving the observables
\begin{equation} \label{eq: gammas 2well}
    \Gamma_1(\mathbf{x}) =  -\frac{\partial (\log{\rho_0})}{\partial x}, \qquad \Gamma_2(\mathbf{x}) =  -\left(4 +4x\frac{\partial (\log{\rho_0})}{\partial x}\right).
\end{equation}
As we have access to the analytic form of the invariant density \eqref{eq: inv dens db well}, we have 
\begin{equation}
    \frac{\partial (\log{\rho_0})}{\partial x}=-\frac{2}{\sigma^2} \frac{\partial V}{\partial x},
\end{equation} 
so we can simplify these expressions:
\begin{equation} \label{eq: gammas 2well explicit}
    \Gamma_1(\mathbf{x}) =  \frac{2}{\sigma^2} \left( 4x^3 - 4x \right), \qquad \Gamma_2(\mathbf{x}) = \frac{32}{\sigma^2} \left( x^4 - x^2 \right) -4.
\end{equation}
These are functions that are already written as a combination of just three of our dictionary functions, so it is straightforward to obtain a decomposition in terms of $\varphi_j(\mathbf{x})$.
The response of the observable $f(x,y) = x$ to the uniform perturbation $\mathbf{X}_1$ is plotted in Figure \ref{fig: GFs 2well} (a), where results of spectral decomposition are seen to agree well with numerical experiments. The Green's function exhibits two distinct timescales: rapid initial decay, followed by a much slower decay to zero. This suggests that $\Gamma_1$ projects strongly on eigenfunctions $\varphi_j$, $j>1$, and so terms in the sum given in Eq. \eqref{eq: spectral Green 1 version} with $k>1$ dominate at early times, with a crossover to slower decay set by $\lambda_1$ at $t \approx 1/|\lambda_2| \approx 0.5$, as clear from the inset in Figure \ref{fig: GFs 2well} (a).
\\
We note that $\Gamma_2$ is an even function of $x$. Hence, any observable that is odd in $x$ will have a vanishing response to the perturbation $\mathbf{X}_2$. 
We consider then response of the observable  $f(x,y) = x^2 + y^2$. 
Again, Figure \ref{fig: GFs 2well} (b) shows good agreement between our spectral decomposition method and direct numerical response experiments. As there is no projection onto the slow-decaying, leading term in \eqref{eq: spectral Green 1 version}, the decay of this Green's function is exponential on a much faster time scale than in the case portrayed in Figure \ref{fig: GFs 2well} (a).
\\
 We tested the robustness of our method by varying the length $T$ of the trajectory used to evaluate the Kolmogorov properties. For each $T$, we performed an ensemble of $60$ independent realisations and tracked the average and standard deviation of our estimate. Figure \ref{fig: GFs 2well} (b) shows that robust and reliable response properties are  well identified with a trajectory as short as $T = 500$. On the contrary, Figure \ref{fig: GFs 2well} (a) shows that longer trajectories are needed to obtain an unbiased estimate of the Green's function, with its standard deviation decreasing, as expected, for longer trajectories. 
 \\
 Such difference in data requirement between panel (a) and (b) in \ref{fig: GFs 2well} is to be expected. Panel (b) refers to observables that are agnostic to the interwell dynamics, that is, they do not distinguish between the two wells. On the contrary, Panel (a) corresponds to the observable $f(x,y) = x$ which is strongly affected by the hopping of the system between the two wells. To properly estimate dynamical properties of such observables the training data should include visits to both wells. Since we are using an ergodic sampling, observing transitions between the wells in the training data is rare and longer trajectories are needed to properly sample the full phase space.  }
\section{Conclusions}\label{conclusions}
Koopmanism provides a powerful theory-informed data-driven method for analysing a large class of dynamical systems, be they deterministic or stochastic, and for deducing from a sufficient amount of sufficiently precise observations the laws that determine the evolution of rather general observables of the system under consideration. Such information can be typically collected by observing one long trajectory or multiple short trajectories, each starting from different initial conditions. These two scenarios correspond to separate experimental situations. 

The evolution is controlled by the Koopman (or Kolmogorov in the case of diffusive stochastic systems) operator. The spectral properties of this linear operator enable one to define the invariant measure describing the long term one-time statistics, as well as to compute the rate of decay of correlations between general observables of the system. We have used this latter property to re-interpret the FDT, which links the response of the system to (possibly time-dependent) forcings with its steady state correlation properties. 

Indeed, by taking into account the eigenvalues and eigenvectors of the unperturbed system, it is possible to express the Green's function describing the response of a system to a forcing as a sum of terms, each associated with a specific mode of variability, thus vastly increasing the interpretability of the FDT, and becoming able to identify feedbacks acting at different spatial and temporal scales. 

We have applied this novel framework to the study of three conceptual mathematical models, namely a deterministic one-dimensional chaotic map, a two-dimensional stochastic map, and a two-dimensional gradient flow, the latter one corresponding to a paradigmatic case of statistical mechanical equilibrium. Using the EDMD algorithm we have computed the eigenvectors and eigenvalues of the Koopman/Kolmogorov operators, which are then used as building blocks for finding explicit expressions for correlation functions of general observables and response operators for given acting forcings and target observables. The reconstructed correlation and Green's functions feature an excellent agreement with those derived by performing direct numerical experiments, thanks to the properties of strong operator convergence property of the EDMD algorithm.  

Thanks to the spectral decomposition, we are able to explore the multiscale nature of the different temporal scales and spatial patterns associated with the various modes of variability of the system and identify under which circumstances specific forcings resonate with the underlying modes of variability of the system. We remind the reader however that there is no guarantee that the output of the EDMD algorithm converges to spectral properties of the actual Koopman/Kolmogorov operator. Hence, one should in general carefully test the physical and mathematical meaningfulness of each term of the decomposition by employing 
ResDMD, a more sophisticated version of the algorithm allowing for rigorous error control \cite{Colbrook_ResDMD,Colbrook_Ayton_Szoke_2023}. 

In each of the cases studied we have performed EDMD taking into account information gathered using only forward trajectory exploring the (ergodic) invariant measure of the system. This corresponds, in statistical mechanical terms, to a specific and fairly realistic way of preparing the system. This, though, comes with some disadvantages in stochastic settings. In particular, our procedure makes it hard to estimate the uncertainty of the obtained estimates of (eigenvalue,eigenvector) pairs. A careful analysis requires investigating dynamical processes associated with higher-order Koopman modes \cite{Colbrook_ResDMD_Stochastics}. Nonetheless, the high quality of the reconstruction we have obtained for correlation and response functions strongly supports the robustness of our analysis. 

The key finding of this paper is the first evidence of how Koopman analysis can be used to interpret and predict the response of a system to perturbations. This has clear implications in many areas of complex system science, and provides a way forward for studying critical behaviour, which is usually associated with the onset of slow decay of correlations and diverging response operators, and for identifying the physical processes responsible for it.

The full deployment of the methodology presented here relies on having complete knowledge of the state of the system, whilst in many cases only partial information can be retrieved. In this case, one should consider response theory for a coarse-grained system. In the case where it is possible to treat the coarse-grained system as approximately Markovian, the protocol proposed here is still valid because coarse-graining amounts, by and large, to projecting out the rapidly decaying Kolmogorov modes and considering the dynamics over relatively long time scales. Linear response is not affected much by the coarse-graining procedure \cite{Koltai2018}.  A natural and algorithmically feasible approach for linking forced and free variability for coarse-grained systems as done here could be provided by using response theory for finite state Markov chains \cite{Lucarini2016MarkovChains,SantosJSP}, which may become practically relevant when analysing complex systems if combined with Markov state modelling  \cite{Pande2010,Husic2019}. The first attempt in this direction has led to  rather promising results \cite{Lucarini2025}.
Koopman analysis, in combination with the use of time-delay measurements, can be used to effectively study systems that are only partially observed \cite{Arbabi2017,Zic2020,Kamb2020,WannerMezic2022}. Adapting the results presented in this paper to such a case by blending time delays with other dictionary functions will be the subject of future investigation. We mention that care should be taken when evaluating Koopman features with time delay observables in stochastic systems \cite{WannerMezic2022}. Response theory for non-Markovian systems presents nontrivial challenges \cite{Hanggi1977,LucariniWouters2017,Giuggioli2019}, despite recent advances in the understanding of the link between Ruelle-Pollicott resonances of full vs. partially observed systems \cite{ChekrounJSPI}. It is helpful to remark that the Koopman operator formalism has proved key to showing how to recast as a simpler multilevel Markovian model   \cite{santos2021}
 the integrodifferential equations resulting from the application of the Mori--Zwanzig projection operator  associated with partial observation \cite{mori_transport_1965,zwanzig_memory_1961}. Hence, the Koopman operator formalism might provide a way ahead for predicting accurately how non-Markovian systems respond to perturbations. 

\subsection*{Acknowledgements}
I.M. was partially supported by the ONR project N00014-21-1-2384. V.L. acknowledges the partial support provided by the Horizon Europe Projects ClimTIP (Grant No. 100018693) and Past2Future (Grant No. 101184070), and by ARIA SCOP-PR01-P003 - Advancing Tipping Point Early Warning AdvanTip. V.L. and J.M. acknowledge the partial support provided by the CROPS RETF project funded by the University of Reading and by the EPSRC project LINK (Grant No. EP/Y026675/1). N.Z. has been supported by the Wallenberg Initiative on Networks and Quantum Information (WINQ).

\bibliographystyle{ieeetr}
\bibliography{biblio}
\end{document}